\documentclass[preprint,12pt]{elsarticle} 

\usepackage{amsmath,amsfonts,amssymb}
\usepackage{graphicx}
\usepackage{multirow}
\usepackage{booktabs}       
\usepackage{url}            
\usepackage{float}
\usepackage{fontawesome}
\usepackage{algorithm}
\usepackage{algpseudocode}
\usepackage[table,xcdraw]{xcolor} 
\usepackage{tcolorbox}
\usepackage{hyperref}       
\usepackage[utf8]{inputenc} 
\usepackage[T1]{fontenc}

\journal{Journal of Systems and Software} 

\title{Can Large Language Models Serve as Data Analysts?\\ A Multi-Agent Assisted Approach for Qualitative Data Analysis}

\author[tampere]{Zeeshan Rasheed}
\ead{zeeshan.rasheed@tuni.fi}

\author[tampere]{Muhammad Waseem}
\ead{muhammad.waseem@tuni.fi}

\author[jyvaskyla]{Aakash Ahmad}

\author[tampere]{Kai-Kristian Kemell}

\author[Bolzano]{Wang Xiaofeng}

\author[usn]{Anh Nguyen Duc}

\author[tampere]{Pekka Abrahamsson}
\ead{pekka.abrahamsson@tuni.fi}

\address[tampere]{Tampere University, Finland}

\address {Faculty of Computing and Communications, Lancaster University Leipzig}

\address {Faculty of Engineering, Free University of Bozen Bolzano}

\address[usn]{University of South Eastern Norway, Norway}

\begin{document}

\begin{frontmatter}


\begin{abstract}

\textbf{Context}:
Manual qualitative data analysis is time-intensive and can compromise validity and replicability, affecting analysis design, implementation, and reporting. Large Language Models (LLMs) enable human-bot collaboration in Software Engineering (SE), but their potential for qualitative data analysis in SE remains largely unexplored.

\textbf{Objective}:
The objective of this study is to design and develop an LLM-based multi-agent system that synergizes human decision support with AI to automate various qualitative data analysis approaches.

\textbf{Methods}:
We used LLM-based multi-agents systems to assist the qualitative data analysis process, deploying 27 agents, each responsible for a specific task, such as text summarization, initial code generation, and extracting themes and patterns. 

\textbf{Results}: The main findings are: (1) the LLM-based multi-agent system accelerates the qualitative data analysis process, (2) the system effectively automates tasks such as text summarization, initial code generation, and theme extraction, and (3) the publicly accessible code facilitates validation and further evaluation.

\textbf{Conclusion}:
The proposed LLM-based multi-agent system automates qualitative data analysis process, creating opportunities for researchers and practitioners. Future improvements focus on enhancing multilingual performance and integrating continuous expert feedback. The source code of proposed system and system details can be found here: https://github.com/GPT-Laboratory/Qualitative-Analysis-with-an-LLM-Based-Agentts

\end{abstract}


\end{frontmatter}

\section{Introduction}
\label{Introduction}
Large Language Models (LLMs) have transformed academia and various other fields by offering advanced capabilities in Natural Language Processing (NLP), automated content generation, and data analysis \cite{hou2023large, bharathi2024analysis, meyer2023chatgpt}. For instance, LLMs like Generative Pre-training Transformer (GPT) and Llama have shown capabilities in understanding, interpreting, and generating human-like text, making them crucial systems across various domains (e.g., sentiment analysis and document summarization) \cite{fan2023automated}. This technology has opened new directions in various domains of Software Engineering (SE) \cite{wang2023software, fan2023large}.
The adoption of LLMs in SE is primarily driven by an innovative approach that transforms many SE tasks into activities involving data analysis and text classification.
LLMs have shown promise in performing various SE tasks, including software development, text classification, and data analysis\cite{radford2018improving, cao2023comprehensive, peng2023impact, finnie2022robots, ahmad2023towards, rasheed2023autonomous}. 

In the domain of qualitative data analysis in empirical SE, LLMs can also play an important role\cite{xiao2023supporting}. LLMs can automate the extraction and interpretation of large volumes of text data \cite{xiao2023supporting}. For instance, the traditional approach to data analysis is highly dependent on human expertise, which consumes more time and human effort. Qualitative data analysis is dependent on primary data in the form of unstructured text or recordings from interviews. Software systems commonly employed for this purpose include MAXQDA \cite{kuckartz2019analyzing}, Nvivo \cite{hilal2013using}, Atlas.ti \cite{smit2002atlas}, Dedoose \cite{salmona2019qualitative}, WebQDA \cite{costa2018webqda} , and QDAMiner \cite{rietz2021cody}. However, these systems require human input, where researchers manually perform coding, interpret text, and make decisions. This process is time-consuming and requires a highly specific skill set and expertise to ensure the rigor of data analysis.

These traditional approaches to data analysis depend on human expertise, including survey instrumentation, statistical analysis, and interpretation. However, with LLMs, there is a paradigm shift towards more automated, intelligent systems \cite{chew2023llm}. Torii \textit{et al}. \cite{torii2024expanding} mentioned, LLMs have the potential to autonomously perform qualitative data analysis, such as analyzing user feedback, software documentation, and development logs to extract valuable insights. By doing so, LLMs enhance the efficiency of data analysis and bring a level of depth to the interpretation of qualitative data, which was previously challenging to achieve \cite{roberts2024artificial}. Several researchers, including Chew \textit{et al}. \cite{chew2023llm}, Dai \textit{et al}. \cite{dai2023llm}, and Xiao \textit{et al}. \cite{xiao2023supporting} explored the potential of LLMs in qualitative data analysis. However, their efforts primarily focus on automating thematic analysis and content analysis approaches to generate initial codes. Despite these advancements, a notable gap persists in the development of LLM-based systems that can autonomously perform all types of qualitative data analysis processes.
Further research and development are required to address this gap and fully utilize the capabilities of LLMs within qualitative research methodologies. 


In this paper, we propose an LLM-based multi-agent system to assist and automate the qualitative data analysis process, addressing challenges in scalability and efficiency. From the various qualitative research methods, we assist the five most popular methods used in SE: content analysis \cite{defranco2017content}, thematic analysis \cite{braun2006using}, narrative analysis \cite{kellam2015narrative}, grounded theory data analysis \cite{glaser2017discovery}, and discourse analysis \cite{potter2004discourse}.
The proposed system analyzes large textual datasets and interview transcripts to autonomously perform qualitative data analysis. It integrates LLMs to help researchers handle diverse datasets and accelerate the analysis process with improved performance. Primary contributions of this study are:

\begin{itemize}
    \item An LLM-based multi-agent system synergises human decision support with AI to automate various qualitative data analysis approaches, including thematic analysis, grounded theory, content analysis, narrative analysis, and discourse analysis.
\end{itemize}

\begin{itemize}
    \item The integration of LLMs based system into the qualitative analysis process allows researchers to handle large and diverse datasets and speed up the process. This reduces reliance on manual processes, making data analysis more efficient and accessible.  
\end{itemize}


\textbf{Implication of research}: The results of this research can provide foundations to researchers who can formulate new hypotheses about the role of LLMs and explore processes, patterns, and methods for LLM-driven data analysis. Software practitioners and data analysts can follow the reported results to experiment with delegating analytical tasks of data analysis to LLMs.

\textbf{Structure of the paper}: The rest of the paper is organized as follows. We review related work in Section \ref{background} and describe the study methodology in Section \ref{Research Method}. The results of this study are presented in Section \ref{results}. A discussion on key findings and their implications is provided in Section \ref{discussion}. The study concludes with future work in Section \ref{Conclusions}.

\section{Background}
\label{background}

In this section, we briefly present the background of the study with a focus on existing research. Section \ref{LLMsinSE} provides an overview of studies concerning LLMs and SE. Section \ref{QDAinSE} provides a background study on qualitative data analysis in SE. Finally, section \ref{LLMinQDA} examines works that have utilized LLMs for qualitative data analysis.

\subsection{Large Language Models in Software Engineering}
\label{LLMsinSE}
In recent years, LLMs have experienced a rapid advancement in various SE applications \cite{feng2023investigating}. 
As Treude \textit{et al}. \cite{treude2023navigating} mentioned, the LLM and SE are interconnected through the application of NLP techniques to various tasks within the software development lifecycle. LLM's language generation capabilities offer valuable assistance and enhancements to SE processes \cite{thiergart2021understanding, hornemalm2023chatgpt}.
The integration of LLMs into SE has marked a major transformation in this field \cite{allamanis2017learning}. LLMs have demonstrated valuable advantages over traditional methods like models guided by domain-specific languages, probabilistic grammars, and basic neural language models. Nowadays, LLMs have been applied to various field of SE. These include data analysis \cite{rae2021scaling}, text classification \cite{chae2023large}, software development \cite{rasheed2023autonomous}, code search \cite{gu2018deep}, unit test case generation \cite{tufano2020unit}, automated program repair \cite{li2022competition}, and many other.  

Hou \textit{et al}. \cite{quin2024b} conducted a systematic review of 229 papers (2017–2023) on the use of LLMs in SE. They examined different LLM types, data processing techniques, performance optimization methods, and SE tasks where LLMs have been successfully applied, highlighting key challenges, research gaps, and future directions.

Zheng \textit{et al}. \cite{zheng2023towards} reviewed the integration of LLMs into SE, categorizing SE tasks into seven types and highlighting application examples, strengths, and limitations. They identified key challenges, including inconsistent LLM performance, limited evaluation of code-centric models, and the need for customized models tailored to specific SE tasks.
Zheng \textit{et al}. \cite{zheng2023survey} examined the effectiveness of LLMs in the context of SE. They reviewed and evaluated 134 studies focused on code LLMs, highlighting the connections between code-specific LLMs and general-purpose LLMs. Furthermore, they conducted an in-depth analysis of how both general and code LLMs perform in various SE tasks, providing a detailed assessment of their capabilities across different sub-tasks.

Shin \textit{et al}. \cite{shin2023prompt} explored the capabilities of GPT-4 by integrating various prompting techniques (such as basic prompts, context learning, and task-specific prompts) to assess its performance on three common SE tasks: code generation, code summarization, and code translation. They compared GPT-4's performance with 18 other fine-tuned LLMs. Additionally, Li \textit{et al}. \cite{li2024approach} demonstrated the effectiveness of fine-tuning and prompt engineering in automating code reviews, highlighting the role of context-rich prompts in improving model accuracy. Marvin \textit{et al}. \cite{marvin2023prompt} highlighted the adaptability of prompt engineering across diverse SE applications, suggesting its potential to optimize resource allocation in LLM-driven workflows. Together, these studies underline the critical importance of combining personalized prompting techniques with fine-tuning to enhance LLM performance in varied SE contexts.

\subsection{Qualitative Data Analysis in Empirical Software Engineering}
\label{QDAinSE}
Qualitative research is a type of research that focuses on collecting and analyzing non-numerical data (e.g., text, video, or audio) to understand concepts, opinions, or experiences \cite{bailey2008first}. As Seaman \textit{et al.} \cite{seaman1999qualitative} mentioned, qualitative methods are adopted and incorporated into SE research, which help researchers produce richer and more informative results. Qualitative data analysis in SE involves systematic examination of non-numerical elements like developer experiences, user feedback, design ideas, and processes efficiency\cite{kilamo2015social}.
Seaman \textit{et al}. \cite{seaman1999qualitative} introduces qualitative methods in empirical SE. In this paper, \textbf{Seaman} proposed that new research methods are important for examining non-technical aspects and that qualitative methods can be adapted and integrated into the design of empirical studies in SE. 
Motivated by inconsistent review standards, Dittrich \textit{et al}. \cite{dittrich2007editorial} proposed common evaluation criteria for qualitative research quality. In this paper, they proposed eight criteria that highlight the importance of clarity in the contribution of qualitative studies. Dittrich \textit{et al} also highlights the need for detailed procedures in qualitative data analysis to improve the understanding and interpretation of results. 

Among the qualitative research methods, content analysis \cite{defranco2017content}, thematic analysis \cite{braun2006using}, narrative analysis \cite{kellam2015narrative}, grounded theory \cite{glaser2017discovery}, and discourse analysis \cite{potter2004discourse} are the most popular in empirical SE researchers \cite{adolph2012reconciling}, \cite{defranco2017content}.
A content analysis extract the meaning of words or concepts and “goes beyond only counting words to analyze language extensively for the purpose of classifying big amounts of text into an effective number of categories with same meanings” \cite{ahmad2011business}. According to Exton \textit{et al}. \cite{exton2004role}, researchers use content analysis to quantify the usage of SE design patterns as a means of communication within empirical SE communities. Thematic analysis is very similar to content analysis. However, it differs in that the themes are typically not quantified \cite{ahuvia2001traditional}. Thematic analysis focuses on finding themes and also creating categories \cite{braun2006using}. 
Narrative analysis is also a qualitative research method used to understand and interpret the stories people tell about their lives and experiences \cite{cortazzi1994narrative}. Catherine \textit{et al}. \cite{rissman1993narrative} provided a overview of narrative methods in social research, cementing its place as a distinct qualitative method.
Grounded theory starts with a question or an existing theory, followed by a thorough examination of the data. The data are analyzed through constant comparative analysis, where they are tagged with codes and grouped into concepts to identify themes \cite{hsieh2005three}. 
Additionally, the data are analyzed during the collection process, with the analyzed data guiding further data collection \cite{cho2014reducing}.
Finally, discourse analysis is also a qualitative research method used to study written or spoken language in its social context, focusing on the ways language constructs meaning, identities, and power relations \cite{gill2000discourse}. Zellig Harris \cite{harris1970discourse} invented the term ``discourse analysis '' in the 1970s, and his studies on face-to-face interaction laid foundational concepts \cite{prince1978discourse}. In the 1980s and 1990s, scholars like Michel Foucault \cite{steinert1983development} and Norman Fairclough \cite{fairclough1992discourse} expanded the field, linking discourse to power structures and social change, which led to the development of critical discourse analysis \cite{fairclough2013critical}.

These qualitative research methods are also used to analyze the quality of SE studies \cite{thomas2008methods}, \cite{kellam2015narrative}. For instance, Stol \textit{et al}. \cite{stol2016grounded} reviewed 16 studies to evaluate the quality of SE studies using grounded theory. They concluded that many papers fail to generate a theory, do not clearly indicate which variant of grounded theory is used, and lack sufficient methodological details for rigorous evaluation. Additionally, the authors provided guidelines on how to conduct and report grounded theory studies effectively.

\begin{table}[]
\caption{Automation qualitative methods using AI techniques}
\resizebox{\textwidth}{!}{%
\begin{tabular}{|l|l|l|l|}
\hline
\textbf{S No} & \textbf{Paper Title}                                                                                                                           & \textbf{Automation Tasks}                                                                                                                         & \textbf{Reference}                        \\ \hline
01            & \begin{tabular}[c]{@{}l@{}}Cody: An AI-based system to semi-automate \\ coding for qualitative research.\end{tabular}                          & \begin{tabular}[c]{@{}l@{}}Automates coding through code rules \\ and supervised ML\end{tabular}                                                  & Rietz \textit{et al}. \cite{rietz2021cody}        \\ \hline
02            & \begin{tabular}[c]{@{}l@{}}LLM-assisted content analysis: using large \\ language models to support deductive coding\end{tabular}              & \begin{tabular}[c]{@{}l@{}}Automate content analysis process with\\ LLM\end{tabular}                                                              & Chew \textit{et al}. \cite{chew2023llm}           \\ \hline
03            & \begin{tabular}[c]{@{}l@{}}LLM-in-the-loop: leveraging large \\ language model for thematic analysis\end{tabular}                              & Automate thematic analysis with LLM                                                                                                               & Dai \textit{et al}. \cite{dai2023llm}          \\ \hline
04            & \begin{tabular}[c]{@{}l@{}}Automated thematic analysis of health inf-\\ ormation technology (hit) related incident \\ reports\end{tabular}     & \begin{tabular}[c]{@{}l@{}}Automate thematic analysis method \\ through NLP and ML\end{tabular}                                                   & Li \textit{et al}. \cite{li2021automated}       \\ \hline
05            & \begin{tabular}[c]{@{}l@{}}Using LLMs for qualitative analysis can \\ introduce serious bias\end{tabular}                                      & \begin{tabular}[c]{@{}l@{}}Identified the bias in LLMs when using \\ for qualitative research\end{tabular}                                        & Ashwin \textit{et al}. \cite{ashwin2023using}      \\ \hline
06            & \begin{tabular}[c]{@{}l@{}}Developing and testing an automated qua\\ litative assistant (AQUA) to support \\ qualitative analysis\end{tabular} & \begin{tabular}[c]{@{}l@{}}AQUA, an AI assistant, offers transparent \\ and reliable coding of qualitative data\end{tabular}                      & Lennon \textit{et al}. \cite{lennon2021developing}  \\ \hline
07            & Using ChatGPT for Thematic Analysis                                                                                                            & \begin{tabular}[c]{@{}l@{}}Perform thematic analysis by using\\ ChatGPT\end{tabular}                                                              & Turobov \textit{et al}. \cite{turobov2024using}      \\ \hline
08            & \begin{tabular}[c]{@{}l@{}}Automation of Qualitative Content \\ Analysis: A Proposal\end{tabular}                                              & \begin{tabular}[c]{@{}l@{}}Semi automate content analysis by using\\ ML techniques. Only automate the coding\end{tabular}                         & Hoxtel \textit{et al}. \cite{hoxtell2019automation} \\ \hline
09            & \begin{tabular}[c]{@{}l@{}}Semi-automated coding for qualitative \\ research: A user-centered inquiry and \\ initial prototypes\end{tabular}   & \begin{tabular}[c]{@{}l@{}}automate the tedious coding process, de-\\ sire transparent systems that extend coding \\ to large datasets\end{tabular} & Marathe \textit{et al}. \cite{marathe2018semi}       \\ \hline
\end{tabular}%
}
\end{table}

\subsection{Large Language Models in Qualitative Research}
\label{LLMinQDA}

In recent studies, LLMs have demonstrated their capability in tasks similar to deductive coding, offering promising alternatives to traditional supervised learning methods \cite{xiao2023supporting}. These models excel in both zero-shot (no examples) and few-shot (few examples) learning scenarios, addressing some of the limitations inherent in conventional approaches \cite{chew2023llm}. Tornberg \cite{tornberg2023chatgpt} observed that zero-shot GPT-4 annotations outperformed both crowd-workers and experts in terms of accuracy and inter-rater reliability when identifying the political affiliation of tweets from U.S. politicians. A study closely aligned with our research by Xiao \textit{et al}. \cite{xiao2023supporting} and Dai \textit{et al}. \cite{dai2023llm} explored the potential of LLMs in deductive coding. However they just automate the thematic analysis and content analysis approaches to generate initial codes. Their findings suggested that prompts based on existing codebooks were more effective than those with exemplar coding decisions, though both approaches fell short of expert coder performance. Building upon these foundations, our work extends the use of LLMs in qualitative research by employing a multi-agent system based on advanced LLMs to automate all types of qualitative data analysis, representing a step forward beyond existing applications in both scope and complexity. There is a lack of single platform that automate all aspects of qualitative analysis. This study aims to bridge this gap by introducing a platform that integrates LLMs with a multi-agent system.
Our approach aims to provide a versatile system that supports qualitative data analysis using LLM-based AI agents.

\section{Research Method}
\label{Research Method}
This research aims to investigate the efficiency of LLMs to assist qualitative data analysis in SE. Our methodology is structured into two phases, each designed to test and analyze the capabilities of LLMs in this context. Below we discuss how our LLM-based multi-agent system performs qualitative analysis tasks.

\subsection{Research Questions}
Considering the objective of our study, we have formulated the following Research Questions (RQs):

\begin{tcolorbox}[colback=gray!2!white,colframe=black!75!black]
\textit{\textbf{RQ1.} How effective are LLM-based multi-agent systems to assist qualitative data analysis?}
\end{tcolorbox}

The main aim of \textbf{RQ1} is to determine whether LLM-based multi-agent systems are capable of assisting in qualitative data analysis. This question was formulated to evaluate the ability of LLM-based agents to accurately interpret and analyze complex textual data without human intervention.


\subsection{System Design}
\label{Development Process}
\begin{figure}[h!]
    \centering
    \includegraphics[width=0.8\textwidth]{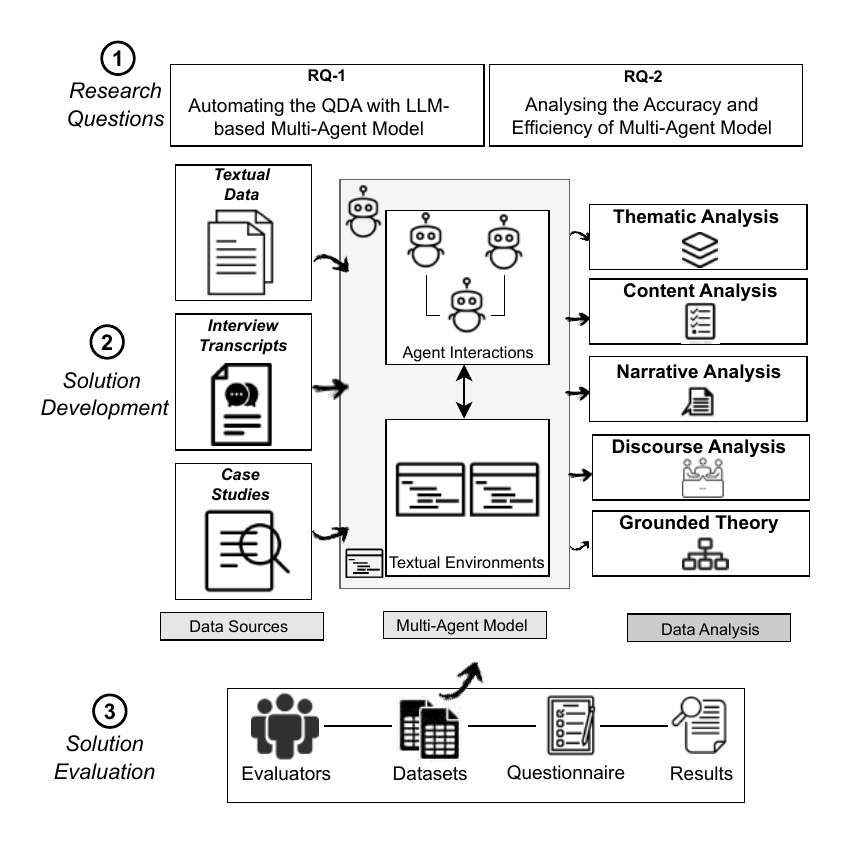}
    \caption{A workflow overview of the proposed system for automation of qualitative data analysis}
    \label{QDA}
\end{figure}

The section reports on how the proposed multi-agent system was developed to assist qualitative data analysis. In this project, we developed 27 agents, each of which is a specialized instance of an LLM, assigned a specific task to perform. As shown in Figure \ref{QDA}, the proposed multi-agent system receives diverse datasets from various sources. The multi-agent system processes the input and sends requests to the OpenAI API environment. Finally, the environment generates and returns a response. This is achieved by utilizing the capabilities of LLMs to understand and interpret complex language structures, making it possible to automate the various aspects of qualitative analysis such as thematic analysis, content analysis, narrative analysis, discourse analysis and grounded theory generation. 
For a more in-depth understanding, we refer to Table \ref{Input output datasource}, showing the diverse datasets employed in our experiment and provides output in multiple formats, including a CSV file, an output area, and documents or PDF files. Below, we discuss the technical details of these agents.

\subsubsection{LLM Based Multi-Agent System}
Firstly, we discuss the technical details of \textit{algorithm 01}, which plays an important role in facilitating communication between agents and users.
As shown in the proposed \textit{algorithm 01}, the process begins by initializing the OpenAI API key. Once the API is initialized, the algorithm reads the input data provided by the user, which can consist of links, textual prompts, or even audio files. After the input is processed, the next step is to submit the input to the selected analysis method. For each selected type, the algorithm initializes conversation histories and assigns modular agents to perform specific tasks. Each agent performs a specific function, such as identifying themes, coding categories, or transcribing narratives, depending on the analysis selected. The agents communicate through a series of interactions with the OpenAI API, processing the input and generating the necessary insights. For audio inputs, transcription and processing are handled using a speech recognition module. Upon completing the analysis, the system parses the results and saves them in the format specified by the user, such as a CSV file or a JSON response. Below, we provide the technical details of each agent.
  



\begin{table}
\scriptsize
\centering
\caption{Input and output data-source}
\label{Input output datasource}
\begin{tabular}{|l|l|l|l|l|}
\hline
\label{Input data and output data}
\textbf{S.No}               & \textbf{QDA Method}                 & \textbf{Input Data Source} & \textbf{Type of Dataset}                                                                             & \textbf{Output Format}      \\ \hline
\multirow{4}{*}{\textbf{1}} & \multirow{4}{*}{Thematic Analysis}  & Links                      & \multirow{4}{*}{\begin{tabular}[c]{@{}l@{}}Github/StackOverflow\\ Developer Discussion\end{tabular}} & CSV File                    \\ \cline{3-3} \cline{5-5} 
                            &                                     & Prompt                     &                                                                                                      & Doc file                    \\ \cline{3-3} \cline{5-5} 
                            &                                     & Upload Files               &                                                                                                      & \multirow{2}{*}{Output Box} \\ \cline{3-3}
                            &                                     & Voice Recording            &                                                                                                      &                             \\ \hline
\multirow{3}{*}{\textbf{2}} & \multirow{3}{*}{Content Analysis}   & Links                      & \multirow{3}{*}{Customer Feedback}                                                                   & CSV File                    \\ \cline{3-3} \cline{5-5} 
                            &                                     & Prompt                     &                                                                                                      & Doc file                    \\ \cline{3-3} \cline{5-5} 
                            &                                     & Upload Files               &                                                                                                      & Output Box                  \\ \hline
\multirow{3}{*}{\textbf{3}} & \multirow{3}{*}{Narrative Analysis} & Links                      & Blogs                                                                                                & CSV File                    \\ \cline{3-5} 
                            &                                     & Prompt                     & \multirow{2}{*}{Human Biography}                                                                     & Doc file                    \\ \cline{3-3} \cline{5-5} 
                            &                                     & Upload Files               &                                                                                                      & Output Box                  \\ \hline
\multirow{3}{*}{\textbf{4}} & \multirow{3}{*}{Discourse Analysis} & Links                      & Customer Reviews                                                                                     & CSV File                    \\ \cline{3-5} 
                            &                                     & Prompt                     & \multirow{2}{*}{Online Conversation}                                                                 & Doc file                    \\ \cline{3-3} \cline{5-5} 
                            &                                     & Upload Files               &                                                                                                      & Output Box                  \\ \hline
\multirow{3}{*}{\textbf{5}} & \multirow{3}{*}{Grounded Theory}    & Links                      & Blogs                                                                                                & CSV File                    \\ \cline{3-5} 
                            &                                     & Prompt                     & \multirow{2}{*}{Observations}                                                                        & Doc file                    \\ \cline{3-3} \cline{5-5} 
                            &                                     & Upload Files               &                                                                                                      & Output Box                  \\ \hline
\end{tabular}
\end{table}

\textbf{\textit{Thematic analysis}} is a method used in qualitative research to identify, analyze, and report patterns (themes) within data \cite{castleberry2018thematic}. It minimally organizes and describes dataset in (rich) detail. It is usually applied to a set of texts. As we can see in Table \ref{thematic analysis}, we developed five agents that work collaboratively to automatically perform thematic analysis on a given input. The \textit{{Agent Summary}} is responsible for receiving the input. This agent primary task is to summarize the text and remove any unnecessary data. The first step is to configure scraper settings adapted to the structure of the discussion from the input data, ensuring the relevant information is captured efficiently. Following this, the algorithm loads and parses the input data to prepare it for summarization. The next step is to generates a clear and concise paragraph that extract the required information from the input data. This summary is then formatted and compiled into a structured format and finally send the summarized data to \textit{Agent Coders} for further action. 
The \textit{{Agent Coders}} receive the summarized data from the \textit{{Agent Summary}} and load the summarized data. The first step is to generate the initial codes to meet thematic requirements. These initial codes pass through a review process and make adjustment to ensure their accuracy and relevance. After this review, the codes are compiled into a structured format. This structured approach ensures that the thematic analysis codes are systematically generated, reviewed, and prepared for further thematic analysis tasks. Finally, the finalized initial thematic codes are then transferred to \textit{{Agent Themes}} for further processing.
The \textit{{Agent Themes}} main tasks is to generate themes for thematic analysis from initial thematic codes provided by \textit{{Agent Coders}}. The process begins by receiving the initial thematic codes from \textit{{Agent Coders}} and organize these codes into preliminary themes, refining them to ensure consistency. After this refinement, the final themes are compiled into a structured format. These documented themes are then transferred to the final report for inclusion in the thematic analysis. This structured approach ensures that thematic codes are systematically organized and refined into meaningful themes that support thematic analysis. The next step is to verify the generated codes and themes. The \textit{{Agent Verify}} received the thematic codes and themes from \textit{{Agent Themes}}.
The process begins by initializing the verification phase, where verification parameters are configured based on quality standards. The themes and initial codes are then received and loaded for verification. The next step involves identifying and resolving any inconsistencies in the themes and codes to ensure accuracy. Once verified, the reviewed data is compiled into a structured format. The verification process and outcomes are documented thoroughly. The verified themes and codes are then transferred to the \textit{{Agent Finalize}} for the final action. 
The \textit{{Agent Finalize}} process starts by receiving the verified themes and codes and configuring the finalization parameters based on storage requirements and data formats. The verified data is then loaded and processed to ensure it matches to the final output standards and formats. The finalized themes and codes are stored in the specified data format in the designated storage system. A final report is then generated, summarizing the themes, codes, and findings. The finalization process concludes with the termination of the process, ensuring that all data is accurately finalized, stored, and documented for future analysis. 

\begin{algorithm}[]
\scriptsize
\caption{Qualitative Data Analysis System with AI Agents}
\begin{algorithmic}[1]
\Require Input data (e.g., project description, links, or audio)
\Ensure Finalized analysis results for each selected type (Thematic, Content, Narrative, Discourse, Grounded Theory)

\State \textbf{Initialize OpenAI API:}
\State Set the \texttt{OpenAI\_API\_KEY} with the appropriate API key.
\State Define the model (e.g., \texttt{gpt-4o}).

\State \textbf{Prepare for Analysis:}
\State Read the \texttt{input\_data} from the user (links, prompts, audio).
\State Identify and select analysis types (Thematic, Content, Narrative, Discourse, Grounded Theory).
\State Initialize \texttt{conversation\_history} for each selected analysis type.

\State \textbf{Call Agents for Analysis:}
\For{each \texttt{selected\_analysis\_type}}

    \If{Thematic Analysis is selected}
        \State Call the following agents:
        \State \texttt{Agent Summary}, \texttt{Agent Coders}, 
        \State \texttt{Agent Themes}, \texttt{Agent Verify}, 
        \State \texttt{Agent Finalize}.
    \ElsIf{Content Analysis is selected}
        \State Call the following agents:
        \State \texttt{Agent Data Pre-processing}, \texttt{Agent Codebook}, 
        \State \texttt{Agent Coding Category}, \texttt{Agent Pattern Recognition}, 
        \State \texttt{Agent Verify}, \texttt{Finalize Agent}.
    \ElsIf{Narrative Analysis is selected}
        \State Call the following agents:
        \State \texttt{Agent Data Transcription}, \texttt{Agent Open Coding}, 
        \State \texttt{Agent Axial Coding}, \texttt{Agent Selective Code}, 
        \State \texttt{Agent Construct Narrative Analysis}, \texttt{Agent Finalize}.
    \ElsIf{Grounded Theory is selected}
        \State Call the following agents:
        \State \texttt{Agent Data Scraper}, \texttt{Agent Coding}, 
        \State \texttt{Agent Focused Coding}, \texttt{Agent Theoretical Coding}, 
        \State \texttt{Agent Theory Development}.
    \ElsIf{Discourse Analysis is selected}
        \State Call the following agents:
        \State \texttt{Data Transcribed Agent}, \texttt{Agent Themes}, 
        \State \texttt{Agent Discourse Analyzer}, \texttt{Agent Contextual Analyzer}, 
        \State \texttt{Agent Finalize}.
    \EndIf

    \State Initialize the conversation history for the selected agent(s).
    \State Call OpenAI API with the agent prompts.
\EndFor

\State \textbf{Agent Interaction:}
\For{each round in \texttt{interaction\_rounds}}
    \State Agent sends a message.
    \State Process the message to extract analysis results.
    \State Update the conversation history with the agent’s response.
\EndFor

\State \textbf{Return Results:}
\State Parse the final agent response.
\State Extract the results for each analysis type.

\State \textbf{Save Processed Data:}
\If{\texttt{output\_type} is CSV}
    \State Save results to a CSV file.
\ElsIf{\texttt{output\_type} is JSON}
    \State Return response as JSON.
\EndIf
\State For audio input, transcribe and process using \texttt{speech\_recognition}.

\end{algorithmic}
\end{algorithm}

\textbf{\textit{{Content analysis} }}is used to analyze media interviews, open-ended surveys, and other forms of text data \cite{drisko2016content}. Researchers use content analysis to track changes over time, compare media content, understand the perspectives of different groups, or identify the prevalence of themes or perspectives in discussions. To automate the process of content analysis, we developed six AI agents. The \textit{{Agent Data Preprocessing}} is the initial step in the autonomous content analysis process.
This agent is tasked with summarizing the story, ensuring it is structured and cleaned for further processing by another agent. The process start with data collection, where the user uploads input data from the client side. The preprocessing phase then summarizes the input data before passing it to the preprocessing module. Next, preprocessing settings are configured by defining parameters and criteria for identifying unnecessary words. The data cleaning phase involves identifying and removing unnecessary words and phrases from the input data. Finally, the cleaned and structured data is submitted to the \textit{{Agent Codebook}}.
The \textit{{Agent Codebook}} transforms structured and cleaned data from the preparation phase into a finalized codebook. The process begins with the identification of initial codes based on the content of the data. These initial codes are then compiled into a single document to form the codebook. The next step involves validating the codebook to ensure consistency and clarity, ensuring that all codes are well-defined and applicable to the data. After validation, the codebook is finalized, making it ready for use. The final step is the transfer of the completed codebook to the \textit{{Agent Coding Category}}, making it available for the coding team to apply in their analysis.


\begin{table}[ht]
\caption{The AI agent workflow to autonomously perform thematic analysis}
\label{thematic analysis}
\resizebox{\textwidth}{!}{%
\begin{tabular}{|l|l|l|}
\hline
\rowcolor[HTML]{C0C0C0} 
\textbf{Algorithm 1 Agent Summary}                                                                                                    & \textbf{Algorithm 2 Agent Coders}                                                                                                                           & \textbf{Algorithm 3 Agent Themes}                                                                                                     \\ \hline
\textbf{Input}: Interview, textual data, case studies                                                                                                                  & \textbf{Input}: Summarized data from Agent Summary                                                                                                                   & \textbf{Input}: Initial thematic codes from Agent Coders                                                                                                               \\
\begin{tabular}[c]{@{}l@{}}\textbf{Output}: Summarized input data and passed to \\ Agent coders\end{tabular}                                                           & \begin{tabular}[c]{@{}l@{}}\textbf{Output}: Initial thematic analysis codes transferred \\ to Agent Themes\end{tabular}                                              & \textbf{Output}: Coherent themes for thematic analysis                                                                                                                 \\
1. Configure Scraper Settings:                                                                                                                                & 1. Initialize Coding:                                                                                                                                       & 1. Initialize Theme Generation:                                                                                                                               \\
\begin{tabular}[c]{@{}l@{}}2. scraper\_settings ← Configure scraper settings \\ based on the discussion structure\end{tabular}                                & \begin{tabular}[c]{@{}l@{}}2. coding\_parameters ← Configure coding parameters \\ based on thematic requirements\end{tabular}                               & \begin{tabular}[c]{@{}l@{}}2. theme\_parameters ← Configure theme generation\\  parameters based on analysis requirements\end{tabular}                        \\
3. content ← Load and parse the input data pages                                                                                                              & \begin{tabular}[c]{@{}l@{}}3. summarized\_data ← Receive and load summarized \\ data from Agent Summary\end{tabular}                                        & \begin{tabular}[c]{@{}l@{}}3. initial\_codes ← Receive and load initial \\ thematic codes from Agent Coders\end{tabular}                                      \\
4. Create Summary:                                                                                                                                            & 4. Generate Initial Codes:                                                                                                                                  & 4. Organize Codes into Themes:                                                                                                                                \\
\begin{tabular}[c]{@{}l@{}}5. summary ← Create a clear and concise paragraph \\ summarizing them\end{tabular}                                                 & \begin{tabular}[c]{@{}l@{}}5. initial\_codes ← Extract and generate initial thematic \\ codes from summarized\_data using coding\_parameters\end{tabular}   & 5. Refine and Consolidate Themes:                                                                                                                             \\
\begin{tabular}[c]{@{}l@{}}6. pass\_to\_agent\_002 ← Compile summaries into \\ a structured format\end{tabular}                                               & \begin{tabular}[c]{@{}l@{}}6. review\_initial\_codes ← Manually review and adjust \\ the initial codes for accuracy and relevance\end{tabular}              & 6. Compile Final Themes:                                                                                                                                      \\
7. \textbf{End Process}                                                                                                                                                & \begin{tabular}[c]{@{}l@{}}7. compiled\_codes ← Compile the reviewed and adjusted \\ initial codes into a structured format\end{tabular}                    & 7. Transfer Themes:                                                                                                                                           \\
                                                                                                                                                              & 8. Transfer Codes:                                                                                                                                          & \begin{tabular}[c]{@{}l@{}}8. send\_to\_final\_report ← Transfer the documented \\ themes for inclusion in the final thematic analysis \\ report\end{tabular} \\
                                                                                                                                                              & \begin{tabular}[c]{@{}l@{}}9. send\_to\_agent\_theme ← Send the compiled initial \\ thematic codes to Agent Themes for further action\end{tabular}          & 9. \textbf{End Process}                                                                                                                                                \\
                                                                                                                                                              & 10. \textbf{End Process}                                                                                                                                             &                                                                                                                                                               \\ \hline
\rowcolor[HTML]{C0C0C0} 
\textbf{Algorithm 4 Agent Verify}                                                                                                                             & \multicolumn{2}{|l|}{\cellcolor[HTML]{C0C0C0}\textbf{Algorithm 5 Agent Finalize}}                                                                                                                                                                                                                                                                 \\ \hline
\textbf{Input}: Themes and initial codes from Agent Themes                                                                                                           & \multicolumn{2}{|l|}{\textbf{Input}: Verified themes and codes from Agent Verify}                                                                                                                                                                                                                                                                          \\
\begin{tabular}[c]{@{}l@{}}\textbf{Output}: Verified themes and codes stored in a specified \\ format\end{tabular}                                                   & \multicolumn{2}{|l|}{\textbf{Output}: Finalized themes and codes stored in a specified format}                                                                                                                                                                                                                                                             \\
1. Initialize Verification:                                                                                                                                 & \multicolumn{2}{|l|}{1. Initialize Finalization:}                                                                                                                                                                                                                                                                                                 \\
\begin{tabular}[c]{@{}l@{}}2. verification\_parameters ← Configure verification \\ parameters based on quality standards\end{tabular}                       & \multicolumn{2}{|l|}{2. finalization\_parameters ← Configure finalization parameters based on storage requirements and data formats}                                                                                                                                                                                                              \\
\begin{tabular}[c]{@{}l@{}}3. themes\_and\_codes ← Receive and load themes \\ and initial codes from Agent Themes\end{tabular}                              & \multicolumn{2}{|l|}{3. verified\_data ← Receive and load verified themes and codes from Agent Verify}                                                                                                                                                                                                                                            \\
\begin{tabular}[c]{@{}l@{}}4. resolve\_discrepancies ← Identify and resolve \\ any discrepancies or inconsistencies in the themes \\ and codes\end{tabular} & \multicolumn{2}{|l|}{4. finalized\_data ← Process verified data to ensure it conforms to final output standards and formats}                                                                                                                                                                                                                      \\
\begin{tabular}[c]{@{}l@{}}5. compiled\_verified\_data ← Compile the verified \\ and reviewed data into a structured format\end{tabular}                    & \multicolumn{2}{|l|}{5. Quality Assurance:}                                                                                                                                                                                                                                                                                                       \\
6. Document Verification:                                                                                                                                   & \multicolumn{2}{|l|}{6. quality\_check ← Perform quality assurance check to ensure data integrity and formatting correctness}                                                                                                                                                                                                                     \\
\begin{tabular}[c]{@{}l@{}}7. document\_verification ← Document the \\ verification process and outcomes\end{tabular}                                       & \multicolumn{2}{|l|}{7. store\_data ← Store the finalized themes and codes in the specified data format in the designated storage system}                                                                                                                                                                                                         \\
8. Transfer Verified Data:                                                                                                                                  & \multicolumn{2}{|l|}{8. Create Report:}                                                                                                                                                                                                                                                                                                           \\
\begin{tabular}[c]{@{}l@{}}9. send\_to\_finalize\_agent ← Transfer the verified \\ themes and codes to Finalize Agent for final action\end{tabular}         & \multicolumn{2}{|l|}{9. create\_final\_report ← Generate a final report summarizing the themes, codes, and findings}                                                                                                                                                                                                                \\
10. \textbf{End Process  }                                                                                                                                           & \multicolumn{2}{|l|}{10. \textbf{End Process}}                                                                                                                                                                                                                                                                                                             \\ \hline
\end{tabular}%
}
\end{table}

The \textit{{Agent Coding Category}} organizes and finalizes a categorized codebook for further use. The workflow begins with the input of a codebook from the \textit{{Agent Codebook}} stage. The first step is to identify code categories based on the content of the codebook, followed by organizing these codes into their respective categories (categorized codes). Once categorized, the agent establishes relationships by describing how different categories relate to each other. These categorized codes are then compiled into structured documents (categorized codebook). Finally, the categorized codebook is ready to be transferred to the \textit{{Agent Pattern Recognition}}.
\textit{{Agent Pattern Recognition}} convert a categorized codebook into a finalized pattern report for next agents. Initially, the agent extracts relevant data points from the categorized codebook. Furthermore, the agent finds initial patterns by extracting codes and analyze the relationship between the codes. Once these relationships are determined, the validation phase ensures the accuracy and relevance of the identified patterns. Detailed descriptions of these relationships and patterns are then documented. The process concludes with the generation of a detail pattern report, which provides a structured summary of the findings. The next step is to validates the pattern report generated by the \textit{{Pattern Recognition Agent}}. Initially, the process involves reviewing the pattern report by documenting any potential issues that arise during the review. Following this, the agent verifies the identified patterns by cross-referencing them with the original data to confirm their validity. In the final step, the relationships identified in the patterns are validated through additional analysis. The process concludes with the distribution of a finalized validation report, which serves as a detail documentation for the next agent. The final step is to finalizes the validation report. The \textit{{Agent Finalize}} reviews the validation feedback and suggests further refinements. Following this, the agent ensures the completeness and accuracy of the report by confirming that all data, patterns, and relationships are correctly represented. In the final step, the agent compiles and stores all the data in the final report.

\textbf{\textit{{Narrative analysis}}} is a qualitative research method used to analyze personal stories, to understand how individuals make sense of events and actions in their lives \cite{earthy2008narrative}. To automate the narrative analysis approach, we developed six AI agents to understand the context and generate the narrative. 
The process starts with receiving the raw data such as audio recording or text. The \textit{{Agent Data Transcription}} prepares data for transcription by confirming the quality of the text and audio to ensure data is enough for transcription. The main task involves transcribing the data, converting the audio into written text. An initial quality check is conducted to remove unnecessary data and correct any evident transcription errors. 

Upon verification, the transcribed text is compiled into a single document and then transferred for further processing. The \textit{{Agent Open Coding}} receive the summarized data. The core task involves applying open codes to the data using inductive coding methods to generate codes directly from the text. These open codes are then documented to form the basis for a draft codebook. The agent proceeds by refining these codes, merging similar ones, and resolving any inconsistencies. This refined set of codes is compiled into an organized and clearly presented open coding report. Finally, the open coding report is transferred to \textit{{Agent Axial coding}} for further steps in narrative analysis, completing this phase of the workflow. This structured approach ensures that raw transcription data is properly coded and documented.

The \textit{{Agent Axial Coding}} begins with the collection of the open coding report generated by the \textit{{Agent Open Coding}}. The initial step involves preparing for axial coding by confirming that the open coding report is properly formatted. The agent then identifies code relationships by determining categories and properties of the codes. Following this, axial codes are developed to describe the nature of these relationships. The final axial coding report ensures that the report clearly presents the axial codes and their relationships. Once completed, this report is transferred to \textit{{Agent Selective Code}} for further analysis. This structured process enables a detailed understanding of the relationships between codes, which is essential for deeper narrative analysis in the following steps.

The \textit{{Agent Selective Coding}} main aim is to generate selective code. The first step is to receive the axial coding report produced by the previous agent. The initial step is to identify core categories by determining the main categories from the axial codes. The agent then generates selective codes that integrate and refine axial codes under these core categories. The process continues with refining the selective codes by merging similar ones to ensure clarity and consistency. The refined selective codes and their relationships are then documented in a selective coding report. Finally, the selective coding report is transferred for use in next stages of analysis. 
\begin{table}[ht]
\caption{The AI agent workflow to autonomously perform content analysis}
\label{content_analysis}
\resizebox{\textwidth}{!}{%
\begin{tabular}{|l|l|l|}
\hline
\rowcolor[HTML]{C0C0C0} 
\textbf{Algorithm 1 Agent Data Preprocessing} & \textbf{Algorithm 2 Agent Codebook} & \textbf{Algorithm 3 Agent Coding Category} \\ \hline
\begin{tabular}[c]{@{}l@{}}\textbf{Input}: User inputs data or uploads data from the client \\ side.\end{tabular} & \textbf{Input}: Summarized data & \begin{tabular}[c]{@{}l@{}}\textbf{Input}: Comprehensive codebook from the agent \\ codebook.\end{tabular} \\ 
\textbf{Output}: Structured and cleaned data for the next agent. & \textbf{Output}: Finalized codebook for the coding team. & \begin{tabular}[c]{@{}l@{}}\textbf{Output}: Finalized categorized codebook for the next \\ agent.\end{tabular} \\ 
01: Collect Data: & 01: Define Initial Codes: & 01: Identify Code Categories: \\ 
02: User inputs data or uploads data from the client side. & \begin{tabular}[c]{@{}l@{}}02: initial\_codes ← Identify initial codes based on the \\ data content and research objectives.\end{tabular} & \begin{tabular}[c]{@{}l@{}}02: categories ← Identify categories based on codebook \\ content.\end{tabular} \\ 
03: Preprocess Data: & 03: Compile Codebook: & \begin{tabular}[c]{@{}l@{}}03: categorized\_codes ← Organize codes into their \\ respective categories.\end{tabular} \\ 
04: summarized\_data ← Summarize the input data. & \begin{tabular}[c]{@{}l@{}}04: codebook ← Compile initial\_codes into a single \\ document\end{tabular} & 04: Establish Relationships: \\ 
\begin{tabular}[c]{@{}l@{}}05: Pass the summarized data to the preprocessing \\ module.\end{tabular} & 05: Validate Codebook: & \begin{tabular}[c]{@{}l@{}}05: relationships ← Describe how different categories\\ relate to each other.\end{tabular} \\ 
06: Configure Preprocessing Settings: & 06: Ensure consistency and clarity in the codebook. & 06: Compile Categorized Codebook \\ 
\begin{tabular}[c]{@{}l@{}}07: settings ← Define parameters for data preproce-\\ ssing.\end{tabular} & 07: Finalize Codebook: & \begin{tabular}[c]{@{}l@{}}07: categorized\_codebook ← Compile categorized\_codes \\ into structure documents\end{tabular} \\ 
\begin{tabular}[c]{@{}l@{}}08: criteria ← Set criteria for identifying unnecessary \\ words.\end{tabular} & 08: Transfere Codebook: & 08: Validate Categorized Codebook: \\ 
09: Clean Data: & \begin{tabular}[c]{@{}l@{}}09: Transfer the codebook to the agent coding category\end{tabular} & \begin{tabular}[c]{@{}l@{}}09: Ensure consistency and clarity in the categorized \\ codebook.\end{tabular} \\ 
\begin{tabular}[c]{@{}l@{}}10: unnecessary\_words ← Identify unnecessary words \\ and phrases from the input data.\end{tabular} & \textbf{End Process} & 10: Finalize Categorized Codebook: \\ 
\begin{tabular}[c]{@{}l@{}}11: cleaned\_data ← Remove unnecessary words and \\ phrases\end{tabular} &  & 11: Transfer Categorized Codebook \\ 
12: Submit to Next Agent: &  & \textbf{End Process} \\ 
\begin{tabular}[c]{@{}l@{}}13: Pass the structured and cleaned data to the next agent \\ for coding.\end{tabular} &  &  \\ 
\textbf{End Process} &  &  \\ \hline
\rowcolor[HTML]{C0C0C0} 
\textbf{Algorithm 4 Agent Pattern Recognition} & \textbf{Algorithm 5 Agent Verify} & \textbf{Algorithm 6 Finalize Agent} \\ \hline
\begin{tabular}[c]{@{}l@{}}\textbf{Input}: Categorized codebook from the code categorization \\ agent.\end{tabular} & \textbf{Input}: Pattern report from the pattern recognition agent. & \begin{tabular}[c]{@{}l@{}}\textbf{Input}: Validation report from the validation and \\ verification agent.\end{tabular} \\ 
\begin{tabular}[c]{@{}l@{}}\textbf{Output}: Finalized pattern report for the next agent.\end{tabular} & \begin{tabular}[c]{@{}l@{}}\textbf{Output}: Finalized validation report for the next agent.\end{tabular} & \begin{tabular}[c]{@{}l@{}}\textbf{Output}: Finalized report for stakeholders and archiving\end{tabular} \\ 
01: Prepare Data for Analysis: & 01: Review Pattern Report: & 01: Review Validation Feedback: \\ 
\begin{tabular}[c]{@{}l@{}}02: analysis\_data ← Extract relevant data points from the \\ categorized codebook\end{tabular} & \begin{tabular}[c]{@{}l@{}}02: review\_notes ← Document any observations or \\ potential issues during the review.\end{tabular} & \begin{tabular}[c]{@{}l@{}}02: review\_notes ← Document any final observations \\ or areas needing further refinement.\end{tabular} \\ 
03: Identify Initial Patterns & 03: Verify Identified Patterns: & 03: Ensure Completeness and Accuracy: \\ 
\begin{tabular}[c]{@{}l@{}}04: initial\_patterns ← Identify frequent co-occurrences and \\ trends in the codes.\end{tabular} & \begin{tabular}[c]{@{}l@{}}04: verified\_patterns ← Confirm the validity of the patterns \\ through cross-referencing with the original data.\end{tabular} & \begin{tabular}[c]{@{}l@{}}04: completeness\_check ← Confirm that all data, patterns, \\ and relationships are correctly represented.\end{tabular} \\ 
05: Analyze Relationships: & 05: Validate Relationships: & 05: Compile Final Report: \\ 
\begin{tabular}[c]{@{}l@{}}06: relationships ← Determine direct and indirect \\ relationships between codes\end{tabular} & \begin{tabular}[c]{@{}l@{}}06: validated\_relationships ← Confirm the relationships \\ through additional analysis or expert validation.\end{tabular} & \begin{tabular}[c]{@{}l@{}}06: final\_report ← Create a comprehensive and polished \\ document ready for dissemination\end{tabular} \\ 
07: Validate Patterns & 07: Distribute Validation Report: & \textbf{End Process} \\ 
\begin{tabular}[c]{@{}l@{}}08: validated\_patterns ← Confirm the validity of identified \\ patterns and clusters\end{tabular} & \textbf{End Process} &  \\ 
\begin{tabular}[c]{@{}l@{}}09: relationship\_documentation ← Create detailed \\ descriptions of relationships and patterns\end{tabular} &  &  \\ 
10: Generate Pattern Report: &  &  \\ 
\begin{tabular}[c]{@{}l@{}}11: pattern\_report ← A structured document summarizing \\ the findings\end{tabular} &  &  \\ 
\textbf{End Process} &  &  \\ \hline
\end{tabular}%
}
\end{table}


The \textit{{Agent Construct Narrative Analysis}} receives the selective coding report produced by the \textit{{Agent Selective Coding}}. The initial step involves preparing for narrative analysis by ensuring the selective coding report is properly formatted. Following this, the agent identifies key themes by determining the main themes that emerge from the data. The next step is to construct narrative segments by developing segments that logically flow and build on each other. The agent then analyzes the relationships and context by describing how different segments and themes relate to each other. This is followed by formulating analytical insights that provide a deeper understanding of the narrative. The process concludes with compiling a narrative analysis report that ensures the report is organized and clearly presents the analyzed narrative. 
The \textit{{Agent Finalize}} workflow start with receiving the final narrative report generated by the \textit{{Agent Construct Narrative Analysis}}. The initial task involves reviewing the final narrative report. Following this, the agent conducts a completeness check to ensure that no essential information is missing, followed by a consistency check to resolve any inconsistencies identified during the review. The next steps involve proofreading and editing to enhance readability and accuracy. Once revisions are complete, the final report is securely stored in an accessible location.

\textbf{\textit{{Ground theory}}} involves in the construction of theories through the methodical gathering and analysis of data \cite{oktay2012grounded}. Grounded theory differs from other qualitative analysis methods as it is not dependent on a pre-existing theory to guide the direction of the research. It allows the researcher to develop a new theory that is grounded in the data that has been collected. To autonomously perform a grounded theory approach on our given input data, we developed five AI agents that work collaboratively to generate theory.

\begin{table}[ht]
\caption{The AI agent workflow to autonomously perform narrative analysis}

\label{narrative_analysis}
\resizebox{\textwidth}{!}{%
\begin{tabular}{|l|l|l|}
\hline
\rowcolor[HTML]{C0C0C0} 
\textbf{Algorithm 01: Agent Data Transcription} & \textbf{Algorithm 02 Agent Open Coding} & \textbf{Algorithm 03 Agent Axial Coding} \\ \hline
\begin{tabular}[c]{@{}l@{}}\textbf{Input}: Raw audio or video recordings from various \\ sources.\end{tabular} & \textbf{Input}: Transcript data & \textbf{Input}: Open coding report \\ 
\begin{tabular}[c]{@{}l@{}}\textbf{Output}: Rough transcription compiled into a single \\ document\end{tabular} & \begin{tabular}[c]{@{}l@{}}\textbf{Output}: Open coding report for further steps in narrative \\ analysis.\end{tabular} & \textbf{Output}: Axial coding report \\ 
01: Collect Raw Data: & 01: Collect Rough Transcription: & 01: Collect Open Coding Report \\ 
02: Prepare for Transcription: & 02: Initial Review of Transcription & 02: Prepare for Axial Coding \\ 
\begin{tabular}[c]{@{}l@{}}03: preparation\_check ← Confirm the quality of the \\ text and audio is adequate for transcription.\end{tabular} & \begin{tabular}[c]{@{}l@{}}03: initial\_notes ← Document initial thoughts and \\ observations.\end{tabular} & \begin{tabular}[c]{@{}l@{}}03: preparation\_check ← Confirm that the open coding \\ report is properly formatted\end{tabular} \\ 
04: Transcribe Data: & 04: Apply Open Codes: & 04: Identify Code Relationships: \\ 
\begin{tabular}[c]{@{}l@{}}05: transcribed\_text ← Convert the audio into written \\ text.\end{tabular} & \begin{tabular}[c]{@{}l@{}}05: open\_codes ← Use inductive coding to generate \\ codes directly\end{tabular} & \begin{tabular}[c]{@{}l@{}}05: relationships\_identified ← Determine categories, \\ properties, and dimensions of codes\end{tabular} \\ 
06: Initial Quality Check: & 06: Document Open Codes & 06: Develop Axial Codes: \\ 
\begin{tabular}[c]{@{}l@{}}07: quality\_check ← Correct any obvious errors or \\ omissions in the transcription.\end{tabular} & \begin{tabular}[c]{@{}l@{}}07: codebook\_draft ← Begin compiling a draft codebook \\ with initial codes\end{tabular} & \begin{tabular}[c]{@{}l@{}}07: axial\_codes ← Generate codes that describe the \\ nature of relationships\end{tabular} \\ 
08: Compile Rough Transcription: & \begin{tabular}[c]{@{}l@{}}08: refined\_codes ← Merge similar codes, refine definitions, \\ and resolve any discrepancies.\end{tabular} & \begin{tabular}[c]{@{}l@{}}08: axial\_codebook ← Compile a codebook with axial \\ codes\end{tabular} \\ 
\begin{tabular}[c]{@{}l@{}}09: rough\_transcription ← Compile the transcribed \\ text into a single document\end{tabular} & \begin{tabular}[c]{@{}l@{}}09: open\_coding\_report ← Ensure the report is organized and \\ clearly presents\end{tabular} & \begin{tabular}[c]{@{}l@{}}09: axial\_coding\_report ← Ensure the report clearly \\ presents the axial codes and the relationships\end{tabular} \\ 
10: Transfer Data & 10: Transfer Open Coding Report & 10: Transfer Axial Coding Report: \\ 
11: \textbf{End Process} & 11: \textbf{End Process} & 11: \textbf{End Process} \\ \hline
\rowcolor[HTML]{C0C0C0} 
\textbf{Algorithm 04 Agent Selective Code} & \textbf{Algorithm 05 Agent Construct Narrative Analysis} & \textbf{Algorithm 06 Agent Finalize} \\ \hline
\textbf{Input}: Axial Coding Report & \textbf{Input}: Selective coding report & \textbf{Input}: Final narrative report \\ 
\textbf{Output}: Selective Coding Report & \textbf{Output}: Construct narrative analysis report & \textbf{Output}: Finalized narrative report \\ 
01: Collect Axial Coding & 01: Collect Selective Coding Report: & 01: Receive Final Narrative Report \\ 
02: Identify Core Categories: & 02: Prepare for Narrative Analysis: & 02: Review Final Narrative Report: \\ 
03: core\_categories ← Determine the main categories & \begin{tabular}[c]{@{}l@{}}03: preparation\_check ← Confirm that the selective \\ coding report is properly formatted\end{tabular} & \begin{tabular}[c]{@{}l@{}}03: review\_notes ← Document any observations or areas \\ needing further refinement.\end{tabular} \\ 
\begin{tabular}[c]{@{}l@{}}04: selective\_codes ← Generate codes that integrate and \\ refine axial codes under core categories.\end{tabular} & 04: Identify Key Themes: & \begin{tabular}[c]{@{}l@{}}04: completeness\_check ← Confirm that no crucial \\ information is missing.\end{tabular} \\ 
\begin{tabular}[c]{@{}l@{}}05: selective\_codebook ← Compile a codebook with \\ selective codes\end{tabular} & \begin{tabular}[c]{@{}l@{}}05: key\_themes ← Determine the main themes that \\ emerge from the data.\end{tabular} & \begin{tabular}[c]{@{}l@{}}05: consistency\_check ← Address any inconsistencies \\ found during the review.\end{tabular} \\ 
\begin{tabular}[c]{@{}l@{}}06: refined\_selective\_codes ← Merge similar selective \\ codes\end{tabular} & 06: Construct Narrative Segments: & 06: Proofread and Edit: \\ 
07: Compile Selective Coding Report: & \begin{tabular}[c]{@{}l@{}}07: narrative\_segments ← Develop segments that \\ logically flow and build on each other.\end{tabular} & \begin{tabular}[c]{@{}l@{}}07: edited\_report ← Make necessary edits to enhance \\ readability and accuracy\end{tabular} \\ 
\begin{tabular}[c]{@{}l@{}}08: selective\_coding\_report ← Ensure the report presents \\ the selective codes and the relationships they describe.\end{tabular} & 08: Analyze Relationships and Context: & \begin{tabular}[c]{@{}l@{}}08: formatted\_report ← Ensure the report is visually \\ appealing and professionally presented\end{tabular} \\ 
09: Transfer Selective Code & \begin{tabular}[c]{@{}l@{}}09: contextual\_analysis ← Describe how different \\ segments and themes relate to each other\end{tabular} & \begin{tabular}[c]{@{}l@{}}09: archive\_location ← Store the report in a secure and \\ easily accessible location.\end{tabular} \\ 
10: \textbf{End Process} & \begin{tabular}[c]{@{}l@{}}10: analytical\_insights ← Formulate insights that \\ provide deeper understanding of the narrative\end{tabular} & 10: \textbf{End Process} \\ 
 & \begin{tabular}[c]{@{}l@{}}11: narrative\_analysis\_report ← Ensure the report is \\ organized and clearly presents the analyzed narrative\end{tabular} &  \\ 
 & 12: Transfer Analytic Report &  \\ 
 & 13: \textbf{End Process} &  \\ \hline
\end{tabular}%
}
\end{table}

The first agent in this process is the \textit{{Agent Data Scraper}}. The workflow of this agent start with the collection of raw data from various sources. The initial step involves preparing for transcription by identifying and documenting all sources of raw data. Following this, the agent cleans the data to ensure it is free from noise and inconsistencies. Once the data is cleaned, it is organized logically and consistently to form structured data. The next step involves compiling this cleaned and structured data into a well organized data report. Finally, the cleaned data report is transferred for further processing to the \textit{{Agent Coding}}. 
\textit{{Agent Coding}} main task is to generate the open coding from given data. These open codes are then organized into broader categories to form a structured coding scheme. The agent then creates a draft codebook that includes all identified codes.
Once the coding and codebook draft are finalized, the coding report is then transferred to \textit{{Agent Focused Coding}} for further processing. This structured approach ensures that the cleaned data is properly coded and organized into meaningful categories.
\textit{{Agent Focused Coding}} workflow start with collecting the coding report produced by the \textit{{Agent Coding}}. The initial step involves identifying codes by selecting those that are most relevant to the topic.
Once the codes are identified, the agent applies focused coding to narrow down the data to its most essential elements. This process involves creating a codebook specifically for the focused codes. The focused coding report is then compiled to ensure that it clearly presents the focused codes.
Finally, the focused coding report is transferred for further analysis to the \textit{{Agent Theoretical Coding}}.
The \textit{{Agent Theoretical Coding}} main role is to identifying the main theoretical constructs from the focused codes.
Once the theoretical constructs are identified, the agent generates codes that describe the relationships between these focused codes and the theoretical constructs. These relationships are then organized into a structured representation, forming the theoretical framework. A codebook is compiled specifically for these theoretical codes, and similar theoretical codes are merged and refined to ensure clarity and consistency.
The agent then prepares a theoretical coding report that clearly presents the theoretical codes and their relationships. Finally, the theoretical coding report is transferred for further processing by subsequent agents.
The \textit{{Agent Theory Development}} first task is to determine the central constructs, referred to as core constructs, from the theoretical codes.
Following this, the agent develops relationships by defining how these core constructs interact with one another. To validate the initial theory, the agent uses examples and data segments as supporting evidence. 
These relationships and constructs are then organized into a theoretical model that clearly illustrates the core constructs and their interactions.
The agent compiles all findings into a theory development report that clearly presents the developed theory. 

\textbf{\textit{{Discourse analysis}}} is a qualitative and interpretive method used in linguistics, social sciences, and anthropology, among other fields, to study the ways in which language is used in context \cite{johnstone2024discourse}. Researchers use this method to analyze the conversation in depth by examining any written or spoken text. In this approach, we used five agents to autonomously perform discourse analysis on the provided data. 

The \textit{{Agent Data Transcribed}} automates the initial stage of discourse analysis by transforming raw input data, such as spoken conversations, media content, and online communications, into a structured textual format. The workflow start by converting audio inputs to text by using speech-to-text APIs, and text inputs are refined for consistency. The agent then cleans the data by removing unnecessary elements, correcting errors, and ensuring terminology consistency. This processed data is summarized and stored in a structured format, making it ready for further analysis by subsequent agents.

\begin{table}[ht]
\caption{The AI agent workflow to autonomously perform ground theory}

\label{ground_theory}
\resizebox{\textwidth}{!}{%
\begin{tabular}{|l|l|l|}
\hline
\rowcolor[HTML]{C0C0C0} 
\textbf{Algorithm 01 Agent Data Scraper} & \textbf{Algorithm 02 Agent Coding} & \textbf{Algorithm 03 Agent Focused Coding} \\ \hline
\textbf{Input}: Raw data from various sources & \textbf{Input}: Cleaned data report from the data scraper agent. & \textbf{Input}: Coding report \\ 
\begin{tabular}[c]{@{}l@{}}\textbf{Output}: Cleaned data report compiled and ready for \\ further processing\end{tabular} & \textbf{Output}: Coding report & \textbf{Output}: Focused coding report \\ 
01: Collect Raw Data: & 01: Collect Cleaned Data: & 01: Collect Coding Report \\ 
02: Prepare for Transcription: & \begin{tabular}[c]{@{}l@{}}02: preparation\_check ← Confirm that the cleaned data \\ report properly\end{tabular} & 02: Identify Significant Codes: \\ 
\begin{tabular}[c]{@{}l@{}}03: sources\_identified ← Identify and document all \\ sources of raw data\end{tabular} & 03: Open Coding: & \begin{tabular}[c]{@{}l@{}}03: significant\_codes ← Select codes that relevant to\\ the topic\end{tabular} \\ 
04: Clean Data: & \begin{tabular}[c]{@{}l@{}}04: open\_codes ← Generate codes directly from the \\ data\end{tabular} & 04: Apply Focused Coding: \\ 
\begin{tabular}[c]{@{}l@{}}05: cleaned\_data ← Ensure that the data is free from \\ noise and inconsistencies.\end{tabular} & \begin{tabular}[c]{@{}l@{}}05: code\_categories ← Organize open codes into \\ broader categories.\end{tabular} & \begin{tabular}[c]{@{}l@{}}05: focused\_codes ← Apply focused coding to narrow \\ down the data to the most essential elements.\end{tabular} \\ 
\begin{tabular}[c]{@{}l@{}}06: structured\_data ← Ensure the data is organized \\ logically and consistently.\end{tabular} & 06: codebook\_draft ← Create a draft codebook with codes & \begin{tabular}[c]{@{}l@{}}06: focused\_codebook ← Create a codebook specifically \\ for the focused codes.\end{tabular} \\ 
07: Compile Data Report: & 07: refined\_codes ← Merge similar codes and refined it & \begin{tabular}[c]{@{}l@{}}07: focused\_coding\_report ← Ensure the report clearly \\ presents the focused codes\end{tabular} \\ 
\begin{tabular}[c]{@{}l@{}}08: data\_report ← Ensure the report is well-organized \\ and cleaned data.\end{tabular} & 08: Save and Backup & \begin{tabular}[c]{@{}l@{}}08: backup\_created ← Create a backup of the report and \\ codebook\end{tabular} \\ 
09: backup\_created ← Create a backup of the data & \begin{tabular}[c]{@{}l@{}}09: backup\_created ← Create a backup of the report and \\ codebook\end{tabular} & 09: Transfer Focused Coding Report: \\ 
10: Transfer Data & 10: Transfer Coding Report & 10: \textbf{End Process} \\ 
11: \textbf{End Process} & 11: \textbf{End Process} &  \\ \hline
\rowcolor[HTML]{C0C0C0} 
\textbf{Algorithm 04 Agent Theoretical Coding} & \multicolumn{2}{l|}{\cellcolor[HTML]{C0C0C0}\textbf{Algorithm 05 Agent Theory Development}} \\ \hline
\textbf{Input}: Focused coding report & \multicolumn{2}{l|}{\textbf{Input}: Theoretical coding report} \\ 
\begin{tabular}[c]{@{}l@{}}\textbf{Output}: Theoretical coding report and theoretical \\ codebook\end{tabular} & \multicolumn{2}{l|}{\textbf{Output}: Theory development report} \\ 
01: Collect Focused Coding & \multicolumn{2}{l|}{01: Collect Theoretical Coding Report:} \\ 
02: Identify Theoretical Constructs: & \multicolumn{2}{l|}{02: core\_constructs ← Determine the central constructs} \\ 
\begin{tabular}[c]{@{}l@{}}03: theoretical\_constructs ← Determine the main \\ theoretical constructs\end{tabular} & \multicolumn{2}{l|}{03: Develop Relationships} \\ 
\begin{tabular}[c]{@{}l@{}}04: theoretical\_codes ← Generate codes that describe \\ the relationships between focused codes and theoretical \\ constructs\end{tabular} & \multicolumn{2}{l|}{\begin{tabular}[c]{@{}l@{}}04: construct\_relationships ← Define how the core constructs \\ interact\end{tabular}} \\ 
\begin{tabular}[c]{@{}l@{}}05: theoretical\_framework ← Create a structured repre-\\ sentation of the theoretical constructs\end{tabular} & \multicolumn{2}{l|}{\begin{tabular}[c]{@{}l@{}}05: supporting\_evidence ← Use examples and data segments to \\ validate the initial theory\end{tabular}} \\ 
\begin{tabular}[c]{@{}l@{}}06: theoretical\_codebook ← Compile a codebook speci-\\ fically for the theoretical codes.\end{tabular} & \multicolumn{2}{l|}{\begin{tabular}[c]{@{}l@{}}06: theoretical\_model ← Ensure the model clearly illustrates the core \\ constructs and their relationships.\end{tabular}} \\ 
\begin{tabular}[c]{@{}l@{}}07: refined\_theoretical\_codes ← Merge similar theoretical \\ codes and refined it\end{tabular} & \multicolumn{2}{l|}{\begin{tabular}[c]{@{}l@{}}07: theory\_development\_report ← Ensure the report clearly \\ presents the developed theory\end{tabular}} \\ 
\begin{tabular}[c]{@{}l@{}}08: theoretical\_coding\_report ← Ensure the report clearly \\ presents the theoretical codes\end{tabular} & \multicolumn{2}{l|}{\begin{tabular}[c]{@{}l@{}}08: backup\_created ← Create a backup of the report and model to \\ prevent data loss.\end{tabular}} \\ 
09: Transfer Theoretical Code & \multicolumn{2}{l|}{09: \textbf{End Process}} \\ 
10: \textbf{End Process} & \multicolumn{2}{l|}{} \\ \hline
\end{tabular}%
}
\end{table}

The \textit{{Agent Themes}} automates theme identification in discourse analysis by processing cleaned and summarized data from the \textit{{Agent Data Transcribed}}. The agent identifies themes by recognizing frequent terms and applying topic modeling techniques. The identified themes are then refined through a review process, including merging or splitting themes. Finally, the refined themes are saved in a structured format and transferred to the \textit{{Agent Discourse Analyser}}, ensuring a robust foundation for subsequent discourse analysis stages.
The \textit{{Agent Discourse Analyser}} automates the analysis of discourse structure, organization, and linguistic features by processing themes identified by the \textit{{Agent Themes}}. The agent take input data from \textit{{Agent Themes}} and analyzes paragraph organization by identifying topic sentences and supporting details, and evaluating sentence complexity and variety. The agent identifies linguistic features such as metaphors, similes, rhetorical devices, and stylistic elements. Word choice and lexical diversity are also analyzed. The results of the discourse analysis are stored in a structured format and transferred to the \textit{{Agent Contextual Analysis}} for further processing.

\begin{table}[ht]
\caption{Table:7 The AI agent workflow to autonomously perform discourse analysis}
\label{discourse_analysis}
\resizebox{\textwidth}{!}{%
\begin{tabular}{|l|l|l|}
\hline
\rowcolor[HTML]{EFEFEF} 
\textbf{Algorithm 01 Data Transcribed Agent} & \textbf{Algorithm 3 Agent Themes} & \textbf{Algorithm 3 Agent Discourse Analyser} \\ \hline
\begin{tabular}[c]{@{}l@{}}\textbf{Input}: Spoken conversation, media content, online \\ communication, etc.\end{tabular} & \begin{tabular}[c]{@{}l@{}}\textbf{Input}: Cleaned and summarized transcribed data from \\ Agent Transcribed.\end{tabular} & \textbf{Input}: Identified themes from Agent Themes. \\ 
\textbf{Output}: Cleaned and summarized transcribed data. & \textbf{Output}: Identified themes from the datasets. & \begin{tabular}[c]{@{}l@{}}\textbf{Output}: Analysis of discourse structure, organization, \\ and linguistic features.\end{tabular} \\ 
1. Collect Input Data: & 1. Collect Cleaned Data: & 1. Collect Thematic Data: \\ 
2. Preprocess Data: & 2. Preprocess Data for Theme Identification: & 2. Receive identified themes from Agent Themes. \\ 
3. extract\_audio(audio\_or\_video\_input) & 3. tokenize\_text(cleaned\_data) & 3. Preprocess Data for Discourse Analysis: \\ 
4. transcriptions ← call\_speech\_to\_text\_api(audio\_data) & 4. pos\_tagging(cleaned\_data) if necessary & 4. Analyze Paragraph Organization: \\ 
5. else if input is text-based then & 5. Identify Themes: & 5. identify\_topic\_sentences\_and\_supporting\_details \\ 
6. transcriptions ← preprocess\_text\_data(text\_data) & 6. frequent\_terms ← identify\_frequent\_terms & 6. evaluate\_sentence\_complexity\_and\_variety(sentences) \\ 
7. Clean and Summarize Data: & 7. potential\_themes ← apply\_topic\_modeling(cleaned\_data) & 7. Perform Sentiment Analysis: \\ 
8. Remove Unnecessary Data: & 8. Refine Themes: & 8. sentiments ← use\_sentiment\_analysis\_systems \\ 
9. identify\_and\_remove\_filler\_words(transcriptions) & 9. review\_themes(themes) & 9. classify\_sentiments\_at\_sentence\_and\_paragraph\_levels \\ 
10. remove\_irrelevant\_sections(transcriptions) & 10. merge\_split\_themes(themes) & 10. Identify Linguistic Features: \\ 
11. correct\_transcription\_errors(transcriptions) & 11. validate\_themes\_with\_experts(themes) if applicable & 11. detect\_metaphors\_and\_similes(sentences) \\ 
12. ensure\_consistency\_in\_terminology(transcriptions) & 12. merge\_split\_themes(themes) & 12. identify\_rhetorical\_devices\_and\_stylistic\_elements \\ 
13. Summarize Data: & 13. validate\_themes\_with\_experts(themes) if applicable & 13. analyze\_word\_choice\_and\_lexical\_diversity \\ 
14. cleaned\_summarized\_data ← summarize\_transcriptions & 14. Store Identified Themes: & 14. Store Analysis Results: \\ 
\begin{tabular}[c]{@{}l@{}}15. save\_transcriptions(cleaned\_summarized\_data, format=\\ "structured\_format")\end{tabular} & 15. save\_themes(themes, format="structured\_format") & \begin{tabular}[c]{@{}l@{}}15. save\_analysis\_results(discourse\_analysis\_results, \\ format="structured\_format")\end{tabular} \\ 
16. Transfer Data to Agent Themes: & 16. Transfer Data to Agent Discourse Analyser & 16. Transfer Data to Agent Contextual Analyser: \\ 
17. \textbf{End Process} & 17. \textbf{End Process} & 17. \textbf{End Process} \\ \hline
\rowcolor[HTML]{EFEFEF} 
\textbf{Algorithm 4 Agent Contextual Analyser} & \multicolumn{2}{l|}{\cellcolor[HTML]{EFEFEF}\textbf{Algorithm 05 Agent Finaliser}} \\ \hline
\textbf{Input}: Analysis of discourse structure. & \multicolumn{2}{l|}{Input: Contextual analysis results from Agent Contextual Analyser.} \\ 
\textbf{Output}: Contextual analysis of the discourse. & \multicolumn{2}{l|}{\textbf{Output}: Finalized data ready for reporting or further use.} \\ 
1. Collect Discourse Analysis Results: & \multicolumn{2}{l|}{1. Collect Contextual Analysis Results:} \\ 
2. Identify Broader Contextual Elements: & \multicolumn{2}{l|}{2. ensure\_completeness\_and\_accuracy(contextual\_analysis\_results)} \\ 
\begin{tabular}[c]{@{}l@{}}3. identify\_relevant\_socio\_cultural\_factors(discourse\_analysis\\ \_results)\end{tabular} & \multicolumn{2}{l|}{3. validate\_coherence\_and\_consistency(contextual\_analysis\_results)} \\ 
4. Situational Context: & \multicolumn{2}{l|}{4. verify\_metadata\_and\_contextual\_information(contextual\_analysis\_results)} \\ 
5. assess\_immediate\_situational\_factors(analysis\_results) & \multicolumn{2}{l|}{5. Summarize Final Insights:} \\ 
\begin{tabular}[c]{@{}l@{}}6. identify\_roles\_and\_relationships\_of\_participants(discourse\\ \_analysis\_results)\end{tabular} & \multicolumn{2}{l|}{6. compile\_key\_findings(contextual\_analysis\_results)} \\ 
7. Integrate Contextual Information: & \multicolumn{2}{l|}{7. highlight\_important\_themes(contextual\_analysis\_results)} \\ 
8. merge\_contextual\_elements\_with\_discourse\_analysis & \multicolumn{2}{l|}{8. prepare\_executive\_summaries(contextual\_analysis\_results)} \\ 
9. highlight\_interplay\_between\_discourse\_and\_context & \multicolumn{2}{l|}{9. Prepare Final Report:} \\ 
10. Perform Contextual Analysis: & \multicolumn{2}{l|}{10. organize\_data\_into\_report(contextual\_analysis\_results)} \\ 
\begin{tabular}[c]{@{}l@{}}11. analyze\_contextual\_influence\_on\_interpretation\end{tabular} & \multicolumn{2}{l|}{11. include\_visual\_aids(contextual\_analysis\_results)} \\ 
\begin{tabular}[c]{@{}l@{}}12. identify\_context\_dependent\_meanings\_and\_implications\end{tabular} & \multicolumn{2}{l|}{12. ensure\_report\_formatting(contextual\_analysis\_results)} \\ 
13. Compile Contextual Analysis Results: & \multicolumn{2}{l|}{13. Store Finalized Data and Report:} \\ 
\begin{tabular}[c]{@{}l@{}}14. summarize\_findings\_for\_each\_contextual\_element \\(contextual\_analysis\_results)\end{tabular} & \multicolumn{2}{l|}{\begin{tabular}[c]{@{}l@{}}14. save\_finalized\_data\_and\_report(contextual\_analysis\_results, format="structured\_format")\end{tabular}} \\ 
\begin{tabular}[c]{@{}l@{}}15. present\_comprehensive\_report\_of\_contextual\_analysis\end{tabular} & \multicolumn{2}{l|}{15. backup\_data\_and\_report(contextual\_analysis\_results)} \\ 
16. Store Contextual Analysis Results: & \multicolumn{2}{l|}{16. \textbf{End Process}} \\ 
\begin{tabular}[c]{@{}l@{}}17. save\_contextual\_analysis\_results(contextual\_analysis\_results, \\ format="structured\_format")\end{tabular} & \multicolumn{2}{l|}{} \\ 
18. Transfer Data to Agent Finaliser: & \multicolumn{2}{l|}{} \\ 
19. \textbf{End Process} & \multicolumn{2}{l|}{} \\ \hline
\end{tabular}%
}
\end{table}

The \textit{{Agent Contextual Analyser}} automates the incorporation of contextual elements into discourse analysis. This agent begins by collecting the results of the discourse analysis and identifying broader contextual elements, such as socio-cultural factors and situational context. It assesses immediate situational factors and identifies roles and relationships of participants to integrate contextual information with the discourse analysis. The agent merges these elements to highlight the interplay between discourse and context, and performs a detailed contextual analysis to understand the influence of context on interpretation, meanings, and implications. The contextual analysis results are compiled, summarized, and presented in a report. These results are then stored in a structured format and transferred to the \textit{{Agent Finaliser}}, ensuring a thorough understanding of the discourse within its broader context.
The \textit{{Agent Finaliser}} is the concluding component in the automated discourse analysis process. It synthesizes and finalizes the contextual analysis results provided by the \textit{{Agent Contextual Analyser}}. It validates the coherence and consistency of the results and verifies metadata and contextual information. The agent then summarizes final insights, highlighting key findings and important themes, and prepares executive summaries. The finalized data and report are stored in a structured format.

\section{Results}
\label{results}
In this section, we present the study results of the proposed LLM-based multi-agent system for qualitative data analysis. Below, we present the results of our LLM-based proposed system in Section \ref{LLM based model RQ1}, specifically reporting the outcomes of RQ1. 

\subsection{Effectiveness of LLM-based Multi Agent System (RQ1)}
\label{LLM based model RQ1}
Our proposed LLM-based multi-agent system assist traditional qualitative data analysis. Our goal is to integrate the LLM-based agents into qualitative data analysis to process and analyze large, diverse datasets and test their effectiveness for qualitative data analysis. The results indicate that integrating LLMs into qualitative research accelerates the analysis process and improves performance. However, certain limitations remain, emphasizing the need for further improvements in their application. Below, we have provided the results of various types of qualitative data analysis.


The proposed LLM-based multi-agent system has been successfully implemented and tested for autonomous qualitative data analysis. First, we demonstrate our proposed system's result for \textbf{Thematic Analysis}. 

\begin{figure}[h!]
    \centering
    \includegraphics[width=1.0\textwidth]{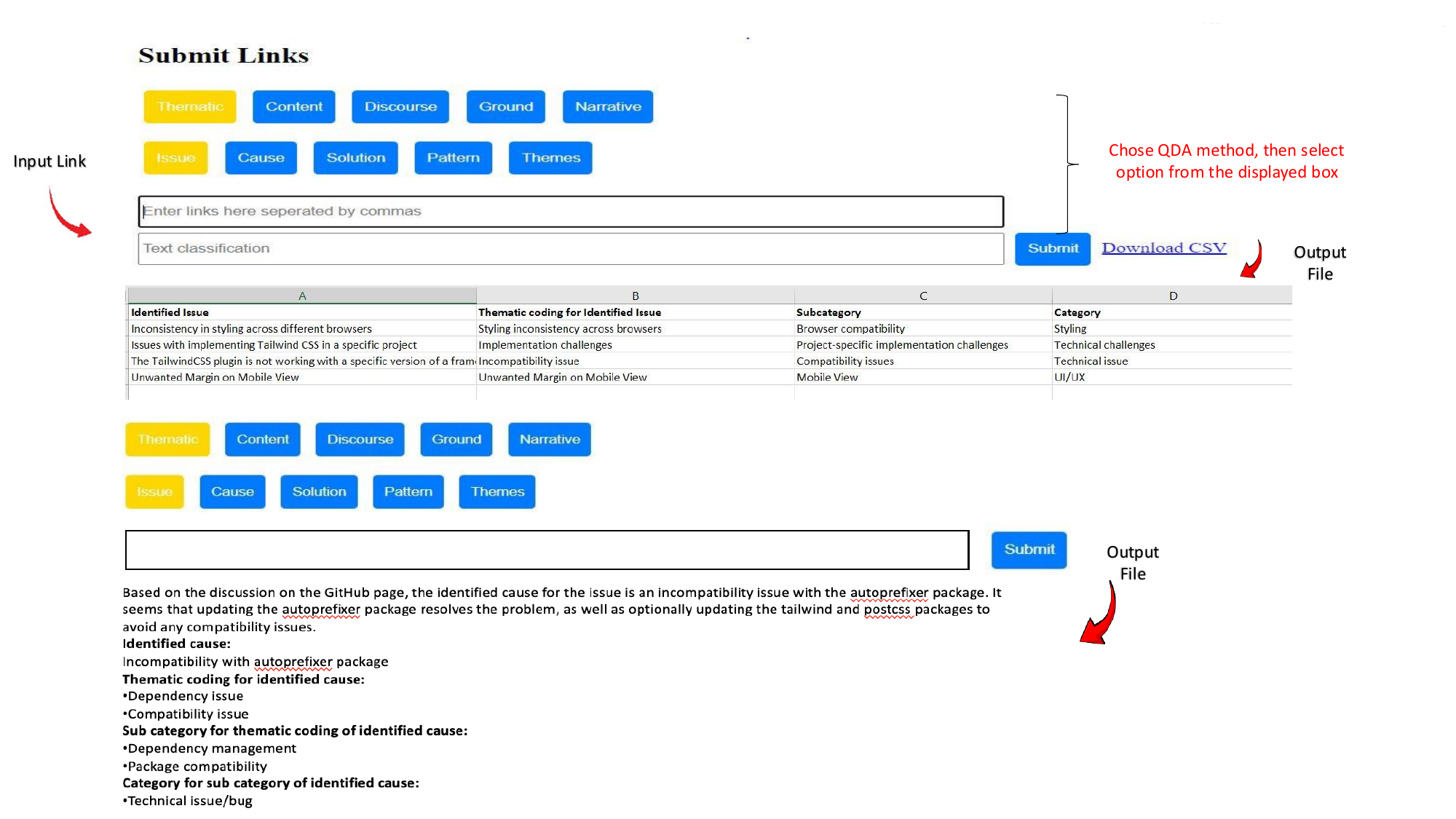}
    \caption{Generated result by proposed system}
    \label{fig:enter-label}
\end{figure}

As shown in Figure \ref{fig:enter-label}, we present a demonstration of
our proposed system for thematic analysis. As depicted, we input GitHub links and select
thematic analysis to prompt the system to identify issues from the
text and perform thematic analysis. The generated output is then
obtained in a CSV file. The system can handle various input formats. Users can choose their
preferred method of input, allowing for a flexible and user-friendly
interaction. Furthermore, we incorporate text extracted from Stack
Overflow into our prompt. Specifically, we opt for thematic analysis
to guide the system in identifying the cause from the provided text.
As illustrated in Figure \ref{fig:enter-label}, the process begins by summarizing the
text and identifying the cause. Subsequently, the system proceeds
to generate initial codes, which are then broken down into subcategories and categories. This time we get final results in output
box. This systematic approach allows for a comprehensive and
structured analysis of the input text. In this project, users have the
autonomy to define and articulate the specific analysis they wish
to perform. This streamlined process allows for efficient and automated exploration of the underlying concepts and themes within
the given data, showcasing the system’s ability to deliver insightful outcomes without manual intervention. Our results demonstrate
the system’s capability to autonomously execute qualitative data
analysis methods on diverse datasets, streamlining the analysis
process and reducing the need for manual intervention.
In our proposed system, we have implemented an additional feature allowing users to upload text, document (doc), and Portable
Document Format (pdf) files as input. The system subsequently autonomously generates responses by activating specific qualitative
analysis approaches, thereby facilitating a comprehensive examination of the provided data. This functionality extends the utility
of our proposed system, making it accessible for practitioners and
researchers alike. Moreover, the system is versatile enough to be
employed for interview analysis, enabling practitioners and researchers to utilize it for conducting interviews and summarizing
the interview text efficiently


Next, we present the results of the proposed system for \textbf{Narrative Analysis}. 
First, we provide the input to our proposed system, which then generates a narrative analysis. The first step is to summarize the text, followed by extracting open coding from the summarized data. In this step, the system breaks down the data into discrete parts. During open coding, the data is labeled with codes that describe what is happening in the text, aiming to uncover key points and their meanings. Second, the system autonomously performs axial coding on the open coding data. Axial coding follows open coding and involves connecting categories to subcategories, linking them through relationships and patterns. This process helps refine and reorganize the initial codes into a coherent structure that shows how different concepts interact with each other. The third step is to autonomously perform selective coding. In this step, the system identifies the core category that integrates all other categories. Finally, the proposed system refines the analysis to form a narrative or theory that best explains the phenomenon under study. By validating the generated response, we observe that the narrative constructs produced by the proposed system lack creativity and holistic storytelling, which a human analyst could provide, making the output feel mechanical.

We now present the results of our proposed system for autonomously performing \textbf{Content Analysis}.  
First the proposed system initially summarizes the given input dataset and then autonomously generates a codebook for the summarized data. Next, the system performs coding categorization based on the initially generated codebook. Finally, the system autonomously generates pattern recognition based on the created categories. The system reduces bias in thematic categorization and pattern recognition, providing an impartial overview of the feedback. On the other hand, the quality of the generated analysis depends heavily on the clarity and consistency of the input data. Poorly structured or ambiguous feedback lead to inaccurate or incomplete insights.

\textbf{Discourse Analysis} focuses on how language constructs and conveys meaning in different contexts. Initially, we provided input and then proposed system first transcribed the data and summarized it for further processing. The next step was to generate themes based on the summarized data. Finally, the proposed system performed a discourse analysis based on the generated themes, helping to understand in detail how these themes are presented, how language is used to shape their meaning and how they resonate with different audiences. LLM-based system effectively extracts and classifies various discourse patterns, such as frustration, satisfaction, fraud, or negligence. However, there is a strong chance that the system may misinterpret context, such as sarcasm, cultural references, or subtle implications, which a human analyst would recognize. 



\textbf{Ground Theory} involves in the construction of theories through the analysis of data. The system first summarized the content to highlight essential information. It then proceeded to perform initial open coding, breaking down the raw data into meaningful codes that represented various concepts and patterns.
Following this, the system perform focused coding, where it break down the open codes to focus on the most relevant and essential elements of the data, refining the analysis. After focused coding, the system moved on to theoretical coding, where it generated relationships between the focused codes and developed constructs, forming a structured theoretical concept that connected the central themes identified during the analysis. Finally, the system produced a theory development report, showing how the core constructs interacted and validating the theory with supporting evidence from the dataset.

\section{Discussion}
\label{discussion}
In this paper, we proposed LLM based multi agent system to assist qualitative analysis process. The multi-agent system interprets a wide range of textual and audio data and autonomously performs various types of qualitative data analysis.

\textbf{LLM-Based Multi-Agent System for Qualitative Analysis (RQ1)}: The results indicate that integrating LLM based multi agents into qualitative analysis is representing a step forward towards automation of big data analysis. Furthermore, we believe that our findings have several implications for the future practice of related research methods. 
Firstly, the implementation of a multi-agent system that interprets vast quantities of textual and audio data highlights the potential of AI in enhancing the efficiency of qualitative research. 
Furthermore, our proposed system opens new opportunities for establishing a new standard in data interpretation and analysis within both industry and academia.
Integrating LLM technology into qualitative analysis accelerates the data analysis process, reduces the manual effort and time required for practitioners and researchers to analyze large datasets, and ensures high accuracy.
This work has broader implications, particularly for large-scale industries. Furthermore, it has the potential to reduce the costs for qualitative studies, as the system is capable of handling complex analytical tasks independently.

However, despite these implications, there are still some challenges that need to be addressed in the future. Firstly, we utilized LLMs to perform qualitative data analysis. However, regarding data privacy, certain ethical concerns were raised, particularly related to the potential exposure of sensitive information and the ethical implications of using LLMs for data analysis. Although we designed specific prompts for each step in qualitative analysis that effectively generated the output to meet our objectives, it is important to acknowledge that we cannot claim these prompts to be the most optimal or accurate. According to Wei \textit{et al}. \cite{wei2022chain}, there is still a gap in exploring the potential of prompts to extract better outputs from LLMs. We believe that improved versions of prompts may lead to better outcomes. Finally, we only utilized OpenAI API, while many others LLM are available. Therefore, we cannot guarantee that using other LLMs will achieve comparable results. Future work could apply this framework to other open-source LLMs, such as LLaMA \cite{touvron2023llama} and Falcon \cite{almazrouei2023falcon}. However, the computational resources and cost might be the potential limitations. We also acknowledge that the proposed system provides autonomous analysis, the use of LLMs poses potential privacy concerns, especially if sensitive qualitative data are processed. The study did not extensively address data governance or privacy protocols, which could be a limitation when deploying the system in practice. We also incorporated a feedback mechanism that enables the system to learn from user input, allowing for continuous improvement in both performance and accuracy.


Our future goal is to involve a more diverse set of participants, including industrial experts, business analysts, and market researchers from various fields where qualitative data analysis plays a critical role. We also plan to utilize benchmarks to broaden the evaluation of our proposed system. Additionally, investigating the model's performance in multilingual settings could broaden its applicability in global research contexts.

\section{Conclusions}
\label{Conclusions}
In this paper, we introduce an LLM-based multi-agent system designed to assist various types of qualitative data analysis, including thematic analysis, content analysis, narrative analysis, discourse analysis, and grounded theory. The main objective is to test the ability of LLM for qualitative data analysis. The initial results of the proposed system indicate that it autonomously performs the analysis on the given dataset. However, there is still a need to highlight the importance of ongoing refinement to address potential areas for improvement.

Our future goal is to focus on exploring the system's performance in multilingual settings to extend its applicability in diverse global research contexts. Additionally, we will continue to emphasize the importance of maintaining a feedback loop with domain experts to ensure the ongoing refinement and enhancement of the system's capabilities. 

\section{Acknowledgment}
This project is co-funded by the European Union and Business Finland under project
BF/Amalia-2023/SW.
\bibliographystyle{unsrtnat}
\bibliography{references}  

\begin{thebibliography}{79}
\providecommand{\natexlab}[1]{#1}
\providecommand{\url}[1]{\texttt{#1}}
\expandafter\ifx\csname urlstyle\endcsname\relax
  \providecommand{\doi}[1]{doi: #1}\else
  \providecommand{\doi}{doi: \begingroup \urlstyle{rm}\Url}\fi

\bibitem[Hou et~al.(2023)Hou, Zhao, Liu, Yang, Wang, Li, Luo, Lo, Grundy, and Wang]{hou2023large}
Xinyi Hou, Yanjie Zhao, Yue Liu, Zhou Yang, Kailong Wang, Li~Li, Xiapu Luo, David Lo, John Grundy, and Haoyu Wang.
\newblock Large language models for software engineering: A systematic literature review.
\newblock \emph{ACM Transactions on Software Engineering and Methodology}, 2023.

\bibitem[Bharathi~Mohan et~al.(2024)Bharathi~Mohan, Prasanna~Kumar, Vishal~Krishh, Keerthinathan, Lavanya, Meghana, Sulthana, and Doss]{bharathi2024analysis}
G~Bharathi~Mohan, R~Prasanna~Kumar, P~Vishal~Krishh, A~Keerthinathan, G~Lavanya, Meka Kavya~Uma Meghana, Sheba Sulthana, and Srinath Doss.
\newblock An analysis of large language models: their impact and potential applications.
\newblock \emph{Knowledge and Information Systems}, pages 1--24, 2024.

\bibitem[Meyer et~al.(2023)Meyer, Urbanowicz, Martin, O’Connor, Li, Peng, Bright, Tatonetti, Won, Gonzalez-Hernandez, et~al.]{meyer2023chatgpt}
Jesse~G Meyer, Ryan~J Urbanowicz, Patrick~CN Martin, Karen O’Connor, Ruowang Li, Pei-Chen Peng, Tiffani~J Bright, Nicholas Tatonetti, Kyoung~Jae Won, Graciela Gonzalez-Hernandez, et~al.
\newblock Chatgpt and large language models in academia: opportunities and challenges.
\newblock \emph{BioData Mining}, 16\penalty0 (1):\penalty0 20, 2023.

\bibitem[Fan et~al.(2023{\natexlab{a}})Fan, Gao, Mirchev, Roychoudhury, and Tan]{fan2023automated}
Zhiyu Fan, Xiang Gao, Martin Mirchev, Abhik Roychoudhury, and Shin~Hwei Tan.
\newblock Automated repair of programs from large language models.
\newblock In \emph{2023 IEEE/ACM 45th International Conference on Software Engineering (ICSE)}, pages 1469--1481. IEEE, 2023{\natexlab{a}}.

\bibitem[Wang et~al.(2023)Wang, Huang, Chen, Liu, Wang, and Wang]{wang2023software}
Junjie Wang, Yuchao Huang, Chunyang Chen, Zhe Liu, Song Wang, and Qing Wang.
\newblock Software testing with large language model: Survey, landscape, and vision.
\newblock \emph{arXiv preprint arXiv:2307.07221}, 2023.

\bibitem[Fan et~al.(2023{\natexlab{b}})Fan, Gokkaya, Harman, Lyubarskiy, Sengupta, Yoo, and Zhang]{fan2023large}
Angela Fan, Beliz Gokkaya, Mark Harman, Mitya Lyubarskiy, Shubho Sengupta, Shin Yoo, and Jie~M Zhang.
\newblock Large language models for software engineering: Survey and open problems.
\newblock In \emph{2023 IEEE/ACM International Conference on Software Engineering: Future of Software Engineering (ICSE-FoSE)}, pages 31--53. IEEE, 2023{\natexlab{b}}.

\bibitem[Radford et~al.(2018)Radford, Narasimhan, Salimans, Sutskever, et~al.]{radford2018improving}
Alec Radford, Karthik Narasimhan, Tim Salimans, Ilya Sutskever, et~al.
\newblock Improving language understanding by generative pre-training.
\newblock 2018.

\bibitem[Cao et~al.(2023)Cao, Li, Liu, Yan, Dai, Yu, and Sun]{cao2023comprehensive}
Yihan Cao, Siyu Li, Yixin Liu, Zhiling Yan, Yutong Dai, Philip~S Yu, and Lichao Sun.
\newblock A comprehensive survey of ai-generated content (aigc): A history of generative ai from gan to chatgpt.
\newblock \emph{arXiv preprint arXiv:2303.04226}, 2023.

\bibitem[Peng et~al.(2023)Peng, Kalliamvakou, Cihon, and Demirer]{peng2023impact}
Sida Peng, Eirini Kalliamvakou, Peter Cihon, and Mert Demirer.
\newblock The impact of ai on developer productivity: Evidence from github copilot.
\newblock \emph{arXiv preprint arXiv:2302.06590}, 2023.

\bibitem[Finnie-Ansley et~al.(2022)Finnie-Ansley, Denny, Becker, Luxton-Reilly, and Prather]{finnie2022robots}
James Finnie-Ansley, Paul Denny, Brett~A Becker, Andrew Luxton-Reilly, and James Prather.
\newblock The robots are coming: Exploring the implications of openai codex on introductory programming.
\newblock In \emph{Proceedings of the 24th Australasian Computing Education Conference}, pages 10--19, 2022.

\bibitem[Ahmad et~al.(2023)Ahmad, Waseem, Liang, Fahmideh, Aktar, and Mikkonen]{ahmad2023towards}
Aakash Ahmad, Muhammad Waseem, Peng Liang, Mahdi Fahmideh, Mst~Shamima Aktar, and Tommi Mikkonen.
\newblock Towards human-bot collaborative software architecting with chatgpt.
\newblock In \emph{Proceedings of the 27th International Conference on Evaluation and Assessment in Software Engineering}, pages 279--285, 2023.

\bibitem[Rasheed et~al.(2023)Rasheed, Waseem, Kemell, Xiaofeng, Duc, Syst{\"a}, and Abrahamsson]{rasheed2023autonomous}
Zeeshan Rasheed, Muhammad Waseem, Kai-Kristian Kemell, Wang Xiaofeng, Anh~Nguyen Duc, Kari Syst{\"a}, and Pekka Abrahamsson.
\newblock Autonomous agents in software development: A vision paper.
\newblock \emph{arXiv preprint arXiv:2311.18440}, 2023.

\bibitem[Xiao et~al.(2023)Xiao, Yuan, Liao, Abdelghani, and Oudeyer]{xiao2023supporting}
Ziang Xiao, Xingdi Yuan, Q~Vera Liao, Rania Abdelghani, and Pierre-Yves Oudeyer.
\newblock Supporting qualitative analysis with large language models: Combining codebook with gpt-3 for deductive coding.
\newblock In \emph{Companion Proceedings of the 28th International Conference on Intelligent User Interfaces}, pages 75--78, 2023.

\bibitem[Kuckartz and R{\"a}diker(2019)]{kuckartz2019analyzing}
Udo Kuckartz and Stefan R{\"a}diker.
\newblock \emph{Analyzing qualitative data with MAXQDA}.
\newblock Springer, 2019.

\bibitem[Hilal and Alabri(2013)]{hilal2013using}
AlYahmady~Hamed Hilal and Saleh~Said Alabri.
\newblock Using nvivo for data analysis in qualitative research.
\newblock \emph{International interdisciplinary journal of education}, 2\penalty0 (2):\penalty0 181--186, 2013.

\bibitem[Smit(2002)]{smit2002atlas}
Brigitte Smit.
\newblock Atlas. ti for qualitative data analysis.
\newblock \emph{Perspectives in education}, 20\penalty0 (3):\penalty0 65--75, 2002.

\bibitem[Salmona et~al.(2019)Salmona, Lieber, and Kaczynski]{salmona2019qualitative}
Michelle Salmona, Eli Lieber, and Dan Kaczynski.
\newblock \emph{Qualitative and mixed methods data analysis using Dedoose: A practical approach for research across the social sciences}.
\newblock Sage Publications, 2019.

\bibitem[Costa et~al.(2018)Costa, de~Souza, Moreira, and de~Souza]{costa2018webqda}
Ant{\'o}nio~Pedro Costa, Francisl{\^e}~Neri de~Souza, Ant{\'o}nio Moreira, and Dayse~Neri de~Souza.
\newblock webqda 2.0 versus webqda 3.0: a comparative study about usability of qualitative data analysis software.
\newblock In \emph{Developments and Advances in Intelligent Systems and Applications}, pages 229--240. Springer, 2018.

\bibitem[Rietz and Maedche(2021)]{rietz2021cody}
Tim Rietz and Alexander Maedche.
\newblock Cody: An ai-based system to semi-automate coding for qualitative research.
\newblock In \emph{Proceedings of the 2021 CHI Conference on Human Factors in Computing Systems}, pages 1--14, 2021.

\bibitem[Chew et~al.(2023)Chew, Bollenbacher, Wenger, Speer, and Kim]{chew2023llm}
Robert Chew, John Bollenbacher, Michael Wenger, Jessica Speer, and Annice Kim.
\newblock Llm-assisted content analysis: Using large language models to support deductive coding.
\newblock \emph{arXiv preprint arXiv:2306.14924}, 2023.

\bibitem[Torii et~al.(2024)Torii, Murakami, and Ochiai]{torii2024expanding}
Maya~Grace Torii, Takahito Murakami, and Yoichi Ochiai.
\newblock Expanding horizons in hci research through llm-driven qualitative analysis.
\newblock \emph{arXiv preprint arXiv:2401.04138}, 2024.

\bibitem[Roberts et~al.(2024)Roberts, Baker, and Andrew]{roberts2024artificial}
John Roberts, Max Baker, and Jane Andrew.
\newblock Artificial intelligence and qualitative research: The promise and perils of large language model (llm)‘assistance’.
\newblock \emph{Critical Perspectives on Accounting}, 99:\penalty0 102722, 2024.

\bibitem[Dai et~al.(2023)Dai, Xiong, and Ku]{dai2023llm}
Shih-Chieh Dai, Aiping Xiong, and Lun-Wei Ku.
\newblock Llm-in-the-loop: Leveraging large language model for thematic analysis.
\newblock \emph{arXiv preprint arXiv:2310.15100}, 2023.

\bibitem[DeFranco and Laplante(2017)]{defranco2017content}
Joanna~F DeFranco and Phillip~A Laplante.
\newblock A content analysis process for qualitative software engineering research.
\newblock \emph{Innovations in Systems and Software Engineering}, 13:\penalty0 129--141, 2017.

\bibitem[Braun and Clarke(2006)]{braun2006using}
Virginia Braun and Victoria Clarke.
\newblock Using thematic analysis in psychology.
\newblock \emph{Qualitative research in psychology}, 3\penalty0 (2):\penalty0 77--101, 2006.

\bibitem[Kellam et~al.(2015)Kellam, Gerow, and Walther]{kellam2015narrative}
Nadia~N Kellam, Karen~Sweeney Gerow, and Joachim Walther.
\newblock Narrative analysis in engineering education research: Exploring ways of constructing narratives to have resonance with the reader and critical research implications.
\newblock In \emph{2015 ASEE Annual Conference \& Exposition}, pages 26--1184, 2015.

\bibitem[Glaser and Strauss(2017)]{glaser2017discovery}
Barney Glaser and Anselm Strauss.
\newblock \emph{Discovery of grounded theory: Strategies for qualitative research}.
\newblock Routledge, 2017.

\bibitem[Potter(2004)]{potter2004discourse}
Jonathan Potter.
\newblock Discourse analysis.
\newblock \emph{Handbook of data analysis}, pages 607--624, 2004.

\bibitem[Feng et~al.(2023)Feng, Vanam, Cherukupally, Zheng, Qiu, and Chen]{feng2023investigating}
Yunhe Feng, Sreecharan Vanam, Manasa Cherukupally, Weijian Zheng, Meikang Qiu, and Haihua Chen.
\newblock Investigating code generation performance of chat-gpt with crowdsourcing social data.
\newblock In \emph{Proceedings of the 47th IEEE Computer Software and Applications Conference}, pages 1--10, 2023.

\bibitem[Treude(2023)]{treude2023navigating}
Christoph Treude.
\newblock Navigating complexity in software engineering: A prototype for comparing gpt-n solutions.
\newblock \emph{arXiv preprint arXiv:2301.12169}, 2023.

\bibitem[Thiergart et~al.(2021)Thiergart, Huber, and {\"U}bellacker]{thiergart2021understanding}
Jonas Thiergart, Stefan Huber, and Thomas {\"U}bellacker.
\newblock Understanding emails and drafting responses--an approach using gpt-3.
\newblock \emph{arXiv preprint arXiv:2102.03062}, 2021.

\bibitem[H{\"o}rnemalm(2023)]{hornemalm2023chatgpt}
Adam H{\"o}rnemalm.
\newblock Chatgpt as a software development tool: The future of development, 2023.

\bibitem[Allamanis et~al.(2017)Allamanis, Brockschmidt, and Khademi]{allamanis2017learning}
Miltiadis Allamanis, Marc Brockschmidt, and Mahmoud Khademi.
\newblock Learning to represent programs with graphs.
\newblock \emph{arXiv preprint arXiv:1711.00740}, 2017.

\bibitem[Rae et~al.(2021)Rae, Borgeaud, Cai, Millican, Hoffmann, Song, Aslanides, Henderson, Ring, Young, et~al.]{rae2021scaling}
Jack~W Rae, Sebastian Borgeaud, Trevor Cai, Katie Millican, Jordan Hoffmann, Francis Song, John Aslanides, Sarah Henderson, Roman Ring, Susannah Young, et~al.
\newblock Scaling language models: Methods, analysis \& insights from training gopher.
\newblock \emph{arXiv preprint arXiv:2112.11446}, 2021.

\bibitem[Chae and Davidson(2023)]{chae2023large}
Youngjin Chae and Thomas Davidson.
\newblock Large language models for text classification: From zero-shot learning to fine-tuning.
\newblock \emph{Open Science Foundation}, 2023.

\bibitem[Gu et~al.(2018)Gu, Zhang, and Kim]{gu2018deep}
Xiaodong Gu, Hongyu Zhang, and Sunghun Kim.
\newblock Deep code search.
\newblock In \emph{Proceedings of the 40th International Conference on Software Engineering}, pages 933--944, 2018.

\bibitem[Tufano et~al.(2020)Tufano, Drain, Svyatkovskiy, Deng, and Sundaresan]{tufano2020unit}
Michele Tufano, Dawn Drain, Alexey Svyatkovskiy, Shao~Kun Deng, and Neel Sundaresan.
\newblock Unit test case generation with transformers and focal context.
\newblock \emph{arXiv preprint arXiv:2009.05617}, 2020.

\bibitem[Li et~al.(2022)Li, Choi, Chung, Kushman, Schrittwieser, Leblond, Eccles, Keeling, Gimeno, Dal~Lago, et~al.]{li2022competition}
Yujia Li, David Choi, Junyoung Chung, Nate Kushman, Julian Schrittwieser, R{\'e}mi Leblond, Tom Eccles, James Keeling, Felix Gimeno, Agustin Dal~Lago, et~al.
\newblock Competition-level code generation with alphacode.
\newblock \emph{Science}, 378\penalty0 (6624):\penalty0 1092--1097, 2022.

\bibitem[Quin et~al.(2024)Quin, Weyns, Galster, and Silva]{quin2024b}
Federico Quin, Danny Weyns, Matthias Galster, and Camila~Costa Silva.
\newblock A/b testing: a systematic literature review.
\newblock \emph{Journal of Systems and Software}, page 112011, 2024.

\bibitem[Zheng et~al.(2023{\natexlab{a}})Zheng, Ning, Chen, Wang, Chen, Guo, and Wang]{zheng2023towards}
Zibin Zheng, Kaiwen Ning, Jiachi Chen, Yanlin Wang, Wenqing Chen, Lianghong Guo, and Weicheng Wang.
\newblock Towards an understanding of large language models in software engineering tasks.
\newblock \emph{arXiv preprint arXiv:2308.11396}, 2023{\natexlab{a}}.

\bibitem[Zheng et~al.(2023{\natexlab{b}})Zheng, Ning, Wang, Zhang, Zheng, Ye, and Chen]{zheng2023survey}
Zibin Zheng, Kaiwen Ning, Yanlin Wang, Jingwen Zhang, Dewu Zheng, Mingxi Ye, and Jiachi Chen.
\newblock A survey of large language models for code: Evolution, benchmarking, and future trends.
\newblock \emph{arXiv preprint arXiv:2311.10372}, 2023{\natexlab{b}}.

\bibitem[Shin et~al.(2023)Shin, Tang, Mohati, Nayebi, Wang, and Hemmati]{shin2023prompt}
Jiho Shin, Clark Tang, Tahmineh Mohati, Maleknaz Nayebi, Song Wang, and Hadi Hemmati.
\newblock Prompt engineering or fine tuning: An empirical assessment of large language models in automated software engineering tasks.
\newblock \emph{arXiv preprint arXiv:2310.10508}, 2023.

\bibitem[Li et~al.(2024)Li, Shi, and Zhang]{li2024approach}
Youjia Li, Jianjun Shi, and Zheng Zhang.
\newblock An approach for rapid source code development based on chatgpt and prompt engineering.
\newblock \emph{IEEE Access}, 2024.

\bibitem[Marvin et~al.(2023)Marvin, Hellen, Jjingo, and Nakatumba-Nabende]{marvin2023prompt}
Ggaliwango Marvin, Nakayiza Hellen, Daudi Jjingo, and Joyce Nakatumba-Nabende.
\newblock Prompt engineering in large language models.
\newblock In \emph{International conference on data intelligence and cognitive informatics}, pages 387--402. Springer, 2023.

\bibitem[Bailey(2008)]{bailey2008first}
Julia Bailey.
\newblock First steps in qualitative data analysis: transcribing.
\newblock \emph{Family practice}, 25\penalty0 (2):\penalty0 127--131, 2008.

\bibitem[Seaman(1999)]{seaman1999qualitative}
Carolyn~B. Seaman.
\newblock Qualitative methods in empirical studies of software engineering.
\newblock \emph{IEEE Transactions on software engineering}, 25\penalty0 (4):\penalty0 557--572, 1999.

\bibitem[Kilamo et~al.(2015)Kilamo, Lepp{\"a}nen, and Mikkonen]{kilamo2015social}
Terhi Kilamo, Marko Lepp{\"a}nen, and Tommi Mikkonen.
\newblock The social developer: now, then, and tomorrow.
\newblock In \emph{Proceedings of the 7th International Workshop on Social Software Engineering}, pages 41--48, 2015.

\bibitem[Dittrich et~al.(2007)Dittrich, John, Singer, and Tessem]{dittrich2007editorial}
Yvonne Dittrich, Michael John, Janice Singer, and Bj{\o}rnar Tessem.
\newblock Editorial for the special issue on qualitative software engineering research.
\newblock \emph{Information and software technology}, 49\penalty0 (6):\penalty0 531--539, 2007.

\bibitem[Adolph et~al.(2012)Adolph, Kruchten, and Hall]{adolph2012reconciling}
Steve Adolph, Philippe Kruchten, and Wendy Hall.
\newblock Reconciling perspectives: A grounded theory of how people manage the process of software development.
\newblock \emph{Journal of Systems and Software}, 85\penalty0 (6):\penalty0 1269--1286, 2012.

\bibitem[Ahmad et~al.(2011)Ahmad, Abd~Razak, Abdullah, Sheik~Osman, Mat~Ali, and Rahmat]{ahmad2011business}
Azizah Ahmad, Rafidah Abd~Razak, Mohd~Syazwan Abdullah, Wan~Rozaini Sheik~Osman, Abdul~Bashah Mat~Ali, and Abdul~Razak Rahmat.
\newblock Business intelligence model for sustainability of the malaysian rural telecenters.
\newblock \emph{Journal of Southeast Asian Research}, 2011, 2011.

\bibitem[Exton(2004)]{exton2004role}
Christopher Exton.
\newblock The role of content analysis in the development of theory and understanding of software engineering.
\newblock In \emph{12 International Workshop on Software Technology and Engineering Practice (STEP'04)}, pages 5--pp. IEEE, 2004.

\bibitem[Ahuvia(2001)]{ahuvia2001traditional}
Aaron Ahuvia.
\newblock Traditional, interpretive, and reception based content analyses: Improving the ability of content analysis to address issues of pragmatic and theoretical concern.
\newblock \emph{Social indicators research}, 54:\penalty0 139--172, 2001.

\bibitem[Cortazzi(1994)]{cortazzi1994narrative}
Martin Cortazzi.
\newblock Narrative analysis.
\newblock \emph{Language teaching}, 27\penalty0 (3):\penalty0 157--170, 1994.

\bibitem[Rissman(1993)]{rissman1993narrative}
Chtherine~Kohler Rissman.
\newblock Narrative analysis: Qualitative research methods, 1993.

\bibitem[Hsieh and Shannon(2005)]{hsieh2005three}
Hsiu-Fang Hsieh and Sarah~E Shannon.
\newblock Three approaches to qualitative content analysis.
\newblock \emph{Qualitative health research}, 15\penalty0 (9):\penalty0 1277--1288, 2005.

\bibitem[Cho and Lee(2014)]{cho2014reducing}
Ji~Young Cho and Eun-Hee Lee.
\newblock Reducing confusion about grounded theory and qualitative content analysis: Similarities and differences.
\newblock \emph{Qualitative report}, 19\penalty0 (32), 2014.

\bibitem[Gill(2000)]{gill2000discourse}
Rosalind Gill.
\newblock Discourse analysis.
\newblock \emph{Qualitative researching with text, image and sound}, 1:\penalty0 172--190, 2000.

\bibitem[Harris and Harris(1970)]{harris1970discourse}
Zellig~S Harris and Zellig~S Harris.
\newblock \emph{Discourse analysis}.
\newblock Springer, 1970.

\bibitem[Prince(1978)]{prince1978discourse}
Ellen~F Prince.
\newblock Discourse analysis in the framework of zellig s. harris.
\newblock \emph{Current Trends in Textlinguistics}, page 191, 1978.

\bibitem[Steinert(1983)]{steinert1983development}
Heinz Steinert.
\newblock The development of" discipline" according to michel foucault: Discourse analysis vs. social history.
\newblock \emph{Crime and Social Justice}, \penalty0 (20):\penalty0 83--98, 1983.

\bibitem[Fairclough(1992)]{fairclough1992discourse}
Norman Fairclough.
\newblock Discourse and text: Linguistic and intertextual analysis within discourse analysis.
\newblock \emph{Discourse \& society}, 3\penalty0 (2):\penalty0 193--217, 1992.

\bibitem[Fairclough(2013)]{fairclough2013critical}
Norman Fairclough.
\newblock Critical discourse analysis.
\newblock In \emph{The Routledge handbook of discourse analysis}, pages 9--20. Routledge, 2013.

\bibitem[Thomas and Harden(2008)]{thomas2008methods}
James Thomas and Angela Harden.
\newblock Methods for the thematic synthesis of qualitative research in systematic reviews.
\newblock \emph{BMC medical research methodology}, 8:\penalty0 1--10, 2008.

\bibitem[Stol et~al.(2016)Stol, Ralph, and Fitzgerald]{stol2016grounded}
Klaas-Jan Stol, Paul Ralph, and Brian Fitzgerald.
\newblock Grounded theory in software engineering research: a critical review and guidelines.
\newblock In \emph{Proceedings of the 38th International conference on software engineering}, pages 120--131, 2016.

\bibitem[Li et~al.(2021)Li, Shyr, Borycki, and Kushniruk]{li2021automated}
Yanyan Li, Casper Shyr, Elizabeth~M Borycki, and Andre~W Kushniruk.
\newblock Automated thematic analysis of health information technology (hit) related incident reports.
\newblock \emph{Knowledge Management \& E-Learning}, 13\penalty0 (4):\penalty0 408, 2021.

\bibitem[Ashwin et~al.(2023)Ashwin, Chhabra, and Rao]{ashwin2023using}
Julian Ashwin, Aditya Chhabra, and Vijayendra Rao.
\newblock Using large language models for qualitative analysis can introduce serious bias.
\newblock \emph{arXiv preprint arXiv:2309.17147}, 2023.

\bibitem[Lennon et~al.(2021)Lennon, Fraleigh, Van~Scoy, Keshaviah, Hu, Snyder, Miller, Calo, Zgierska, and Griffin]{lennon2021developing}
Robert~P Lennon, Robbie Fraleigh, Lauren~J Van~Scoy, Aparna Keshaviah, Xindi~C Hu, Bethany~L Snyder, Erin~L Miller, William~A Calo, Aleksandra~E Zgierska, and Christopher Griffin.
\newblock Developing and testing an automated qualitative assistant (aqua) to support qualitative analysis.
\newblock \emph{Family medicine and community health}, 9\penalty0 (Suppl 1), 2021.

\bibitem[Turobov et~al.(2024)Turobov, Coyle, and Harding]{turobov2024using}
Aleksei Turobov, Diane Coyle, and Verity Harding.
\newblock Using chatgpt for thematic analysis.
\newblock \emph{arXiv preprint arXiv:2405.08828}, 2024.

\bibitem[Hoxtell(2019)]{hoxtell2019automation}
Annette Hoxtell.
\newblock Automation of qualitative content analysis: A proposal.
\newblock In \emph{Forum: Qualitative Social Research}, volume~20. Freie Universit{\"a}t Berlin, 2019.

\bibitem[Marathe and Toyama(2018)]{marathe2018semi}
Megh Marathe and Kentaro Toyama.
\newblock Semi-automated coding for qualitative research: A user-centered inquiry and initial prototypes.
\newblock In \emph{Proceedings of the 2018 CHI conference on human factors in computing systems}, pages 1--12, 2018.

\bibitem[T{\"o}rnberg(2023)]{tornberg2023chatgpt}
Petter T{\"o}rnberg.
\newblock Chatgpt-4 outperforms experts and crowd workers in annotating political twitter messages with zero-shot learning.
\newblock \emph{arXiv preprint arXiv:2304.06588}, 2023.

\bibitem[Castleberry and Nolen(2018)]{castleberry2018thematic}
Ashley Castleberry and Amanda Nolen.
\newblock Thematic analysis of qualitative research data: Is it as easy as it sounds?
\newblock \emph{Currents in pharmacy teaching and learning}, 10\penalty0 (6):\penalty0 807--815, 2018.

\bibitem[Drisko and Maschi(2016)]{drisko2016content}
James~W Drisko and Tina Maschi.
\newblock \emph{Content analysis}.
\newblock Pocket Guide to Social Work Re, 2016.

\bibitem[Earthy and Cronin(2008)]{earthy2008narrative}
Sarah Earthy and Ann Cronin.
\newblock Narrative analysis.
\newblock In \emph{Researching social life}. Sage, 2008.

\bibitem[Oktay(2012)]{oktay2012grounded}
Julianne~S Oktay.
\newblock \emph{Grounded theory}.
\newblock Pocket Guide to Social Work Re, 2012.

\bibitem[Johnstone and Andrus(2024)]{johnstone2024discourse}
Barbara Johnstone and Jennifer Andrus.
\newblock \emph{Discourse analysis}.
\newblock John Wiley \& Sons, 2024.

\bibitem[Wei et~al.(2022)Wei, Wang, Schuurmans, Bosma, Xia, Chi, Le, Zhou, et~al.]{wei2022chain}
Jason Wei, Xuezhi Wang, Dale Schuurmans, Maarten Bosma, Fei Xia, Ed~Chi, Quoc~V Le, Denny Zhou, et~al.
\newblock Chain-of-thought prompting elicits reasoning in large language models.
\newblock \emph{Advances in neural information processing systems}, 35:\penalty0 24824--24837, 2022.

\bibitem[Touvron et~al.(2023)Touvron, Lavril, Izacard, Martinet, Lachaux, Lacroix, Rozi{\`e}re, Goyal, Hambro, Azhar, et~al.]{touvron2023llama}
Hugo Touvron, Thibaut Lavril, Gautier Izacard, Xavier Martinet, Marie-Anne Lachaux, Timoth{\'e}e Lacroix, Baptiste Rozi{\`e}re, Naman Goyal, Eric Hambro, Faisal Azhar, et~al.
\newblock Llama: Open and efficient foundation language models.
\newblock \emph{arXiv preprint arXiv:2302.13971}, 2023.

\bibitem[Almazrouei et~al.(2023)Almazrouei, Alobeidli, Alshamsi, Cappelli, Cojocaru, Debbah, Goffinet, Heslow, Launay, Malartic, et~al.]{almazrouei2023falcon}
Ebtesam Almazrouei, Hamza Alobeidli, Abdulaziz Alshamsi, Alessandro Cappelli, Ruxandra Cojocaru, Merouane Debbah, Etienne Goffinet, Daniel Heslow, Julien Launay, Quentin Malartic, et~al.
\newblock Falcon-40b: an open large language model with state-of-the-art performance, 2023.

\end{thebibliography}






\end{document}